\documentclass[conference]{IEEEtran}
\usepackage[noadjust]{cite}
\usepackage{amsmath,amssymb,amsfonts}
\usepackage{algorithmic}
\usepackage{graphicx}
\usepackage{textcomp}
\usepackage{xcolor}
\usepackage{rotating}
\usepackage{subcaption}
\usepackage{booktabs}
\usepackage{multirow}
\usepackage{tikz}
\usepackage[braket,qm]{qcircuit}
\usepackage[hidelinks]{hyperref}
\usepackage{pdflscape}
\usepackage{bbm}
\usepackage{bm}
\def\BibTeX{{\rm B\kern-.05em{\sc i\kern-.025em b}\kern-.08em
    T\kern-.1667em\lower.7ex\hbox{E}\kern-.125emX}}

\newenvironment{rotatepage}%
    {\clearpage\pagebreak[4]\global\pdfpageattr\expandafter{\the\pdfpageattr/Rotate 90}}%
    {\clearpage\pagebreak[4]\global\pdfpageattr\expandafter{\the\pdfpageattr/Rotate 0}}%

\newcommand*\circled[1]{\tikz[baseline=(char.base)]{
            \node[shape=circle,draw,inner sep=2pt] (char) {#1};}}

\begin{document}

\title{Quantum Image Representation Methods Using Qutrits}

\author{\IEEEauthorblockN{Ankit Khandelwal}
\IEEEauthorblockA{\textit{TCS Research} \\
\textit{Tata Consultancy Services}\\
Bengaluru, India \\
ankit27.kh@gmail.com}
\and
\IEEEauthorblockN{M Girish Chandra}
\IEEEauthorblockA{\textit{TCS Research} \\
\textit{Tata Consultancy Services}\\
Bengaluru, India \\
m.gchandra@tcs.com}
}

\maketitle

\begin{abstract}
Quantum Image Processing is a recent highlight in the quantum computing field. All previous methods for representing the images as quantum states were defined using qubits. One Quantum Image Representation (QIR) method using qutrits is present in the literature. Inspired by the qubit methods and the higher state-space available for qutrits, multiple QIR methods using qutrits are worked out in this paper. The ternary quantum gates required for the representations are described, and then the implementation details for five qutrit-based QIR methods are given. All the methods have been simulated in software, and example circuits are provided.
\end{abstract}


\section{Introduction}
Quantum Image Processing research is in a nascent stage. Several qubit QIR methods are available in the literature. A review summarising many of them is also available \cite{9268129}. In particular, FRQI \cite{Le2011}, FRQCI \cite{doi:10.1142/S0219749918500053}, MCQI \cite{Sun2013AnRM} and QRCI \cite{WANG2019147} methods are considered to provide their qutrit extensions in this paper. To the best of our knowledge, these extensions are novel. An approach to extend the NEQR \cite{Zhang2013} method was previously followed to define the QTRQ \cite{Dong2022} method.

It has been well known for a long time that it is, in principle, possible to use qudits with more than two states  to carry out Quantum Information Processing \cite{Kues2017}.
Compared to qubit, qudit systems provide a more extensive state-space to store and process information and can provide certain advantages such as reduction of the circuit depth/complexity \cite{Lanyon2009,10.3389/fphy.2020.589504,PhysRevA.75.022313}.
In the present work, we are restricting our focus to qutrit-based computing. It is useful to note that there have been physical implementations of qutrit systems \cite{PhysRevLett.125.180504,PhysRevX.11.021010}; Rigetti\footnote{\href{https://medium.com/rigetti/beyond-qubits-unlocking-the-third-state-in-quantum-processors-12d2f84133c4}{https://medium.com/rigetti/beyond-qubits-unlocking-the-third-state-in-quantum-processors-12d2f84133c4}} has also announced experimental access to qutrits.
While Cirq \cite{cirq_developers_2022_6599601} already provides qudit simulation functionality, it is also under development at PennyLane \cite{https://doi.org/10.48550/arxiv.1811.04968}.

The paper is organised as follows. The qutrit quantum gates used in implementing the representations are defined in Section \ref{gates}. The steps to implement the new qutrit-based QIR methods are defined and described in Section \ref{qir}. Conclusion and pointers for future work are provided in Section \ref{conclude}.

\section{Qutrit Gates}
\label{gates}
The qutrit gates used in the implementations are described in this section.

\subsection{Single Qutrit Ternary Gates}
A corresponding qutrit gate $A^{(  i\,j)}$ can be defined for any qubit gate $A$ by defining $A^{(  i\,j)}$ such that the action of $A$ is on $(i,j)$ basis states only $(i,j \in\{0,1,2\}, i\ne j)$ \cite{https://doi.org/10.48550/arxiv.1105.5485}. This allows us to define three $X$ gates for qutrits:
\begin{align}
    X^{(01)}&=\begin{pmatrix}
0 & 1 & 0\\
1 & 0 & 0\\
0 & 0 & 1
\end{pmatrix}\\
    X^{(02)}&=\begin{pmatrix}
0 & 0 & 1\\
0 & 1 & 0\\
1 & 0 & 0
\end{pmatrix}\\
    X^{(12)}&=\begin{pmatrix}
1 & 0 & 0\\
0 & 0 & 1\\
0 & 1 & 0
\end{pmatrix}
\end{align}
By combining the action of these gates, two more single qutrit gates can be constructed:

\subsubsection{Single-Shift Gate}
The action of this gate on a basis state is given by:
\begin{equation}
    \left[+1\right]\left|x\right>=\left|(x+1)\bmod3\right>
\end{equation}
The gate is defined as:
\begin{equation}
    \left[+1\right]=X^{(01)}X^{(12)}
\end{equation}

\subsubsection{Double-Shift Gate}
The action of this gate on a basis state is given by:
\begin{equation}
    \left[+2\right]\left|x\right>=\left|(x+2)\bmod3\right>
\end{equation}
The gate is defined as:
\begin{equation}
    \left[+2\right]=X^{(12)}X^{(01)}
\end{equation}
This construction can be used to define Hadamard $(H^{(  i\,j)})$ gates too. But, these versions do not create an equal superposition state.

\subsubsection{Ternary Hadamard Gate}
The qutrit extension of the qubit Hadamard gate as defined in \cite{Dong2022} is considered in this paper.
\begin{equation}
    H=\frac{1}{\sqrt{3}}\begin{pmatrix}
1 & 1 & 1\\
1 & e^{\frac{2\pi i}{3}} & e^{\frac{4\pi i}{3}}\\
1 & e^{\frac{4\pi i}{3}} & e^{\frac{8\pi i}{3}}
\end{pmatrix}
\end{equation}
The action of the Hadamard gate on the basis states is given by:
\begin{align}
    H\left|0\right>&=\frac{1}{\sqrt{3}}\left(\left|0\right>+\left|1\right>+\left|2\right>\right),\\
    H\left|1\right>&=\frac{1}{\sqrt{3}}\left(\left|0\right>+e^{\frac{2\pi i}{3}}\left|1\right>+e^{\frac{4\pi i}{3}}\left|2\right>\right),\\
    H\left|2\right>&=\frac{1}{\sqrt{3}}\left(\left|0\right>+e^{\frac{4\pi i}{3}}\left|1\right>+e^{\frac{8\pi i}{3}}\left|2\right>\right),
\end{align}
This version of the qutrit Hadamard gate can thus be used on the $\left|0\right>$ state to get to an equal superposition of the three basis states without any phase.

\subsubsection{Ternary Rotation Gates}
The definition of elementary one-qutrit gates as given in \cite{https://doi.org/10.48550/arxiv.1105.5485} is considered.
\begin{align}
    R_\alpha^{(  i\,j)}\left(\theta\right)&=\exp{\left({\frac{-i}{2}\theta\sigma_\alpha^{( j\,k)}}\right)}, j<k, \alpha \in\{x,y,z\}\\
    \sigma_z^{( j\,k)}&=\left|j\right>\left<j\right|   -\left|k\right>\left<k\right|\\
    \sigma_x^{( j\,k)}&=\left|j\right>\left<k\right|   +\left|k\right>\left<j\right|\\
    \sigma_y^{( j\,k)}&=-i\left|j\right>\left<k\right|  +i\left|k\right>\left<j\right|
\end{align}
Note that these rotation gates can also be obtained by using the previous construction.

The construction can also be used to get extensions of general single-qubit rotation gate $U(\theta,\phi,\delta)$. For example:
\begin{equation}
    U^{(12)}(\theta,\phi,\delta) = \begin{pmatrix}
1&0&0\\
0 & \cos{\frac{\theta}{2}}  & -e^{i\delta}\sin{\frac{\theta}{2}}\\
0&e^{i\phi}\sin{\frac{\theta}{2}} &  e^{i(\delta+\phi)}\cos{\frac{\theta}{2}}
\end{pmatrix}
\end{equation}

\subsubsection{Ternary Identity Gate}
The identity gate for qutrits is the 3D identity matrix, 
\begin{equation}
I=\begin{pmatrix}
1 & 0 & 0\\
0 & 1 & 0\\
0 & 0 & 1
\end{pmatrix}
\end{equation}
Its action on a state $\ket{\psi}$ is given as:
\begin{equation}
    I\ket{\psi}=\ket{\psi}
\end{equation}

\subsection{Controlled Qutrit Gates}
Control qutrits for ternary gates can be in any of the three basis states. Thus, for each single-qutrit gate, three different controlled variations of two-qutrit gates (the control qutrit can be in any of $\ket{0},\ket{1}$ or $\ket{2}$ state), nine different three qutrit gates (the control qutrits can be in any of $\ket{00},\ket{01},\ket{02},\ket{10},\ket{11},\ket{12},\ket{20},\ket{21}$ or $\ket{22}$ state) and so on are possible. The decomposition of any such gate can be assured using the universality of qudit logic \cite{10.3389/fphy.2020.589504}.

\section{Qutrit Quantum Image Representation Methods}
\label{qir}
Some QIR methods using qutrits are described in this section. Cirq has been used to simulate these methods. Cirq provides basic functionality to use qudits by allowing the use of custom quantum gates that can be defined for any dimension qudit. The gates used in the paper are \textit{not} natively available and were defined with the definitions given in the previous section before use, another small contribution to the area through this work. The images considered for encoding are of dimension $3^n\times3^n$.

\begin{figure}[!t]
\centering
\begin{subfigure}{0.08\textheight}
         \centering
         \includegraphics[width=\textwidth]{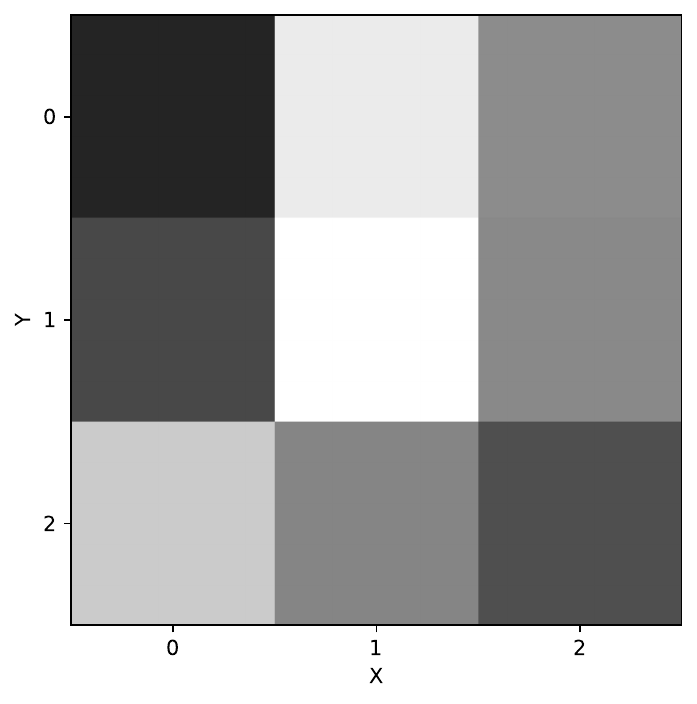}
         \caption{Grayscale}
         \label{fig:gray}
     \end{subfigure}
     \begin{subfigure}{0.08\textheight}
         \centering
         \includegraphics[width=\textwidth]{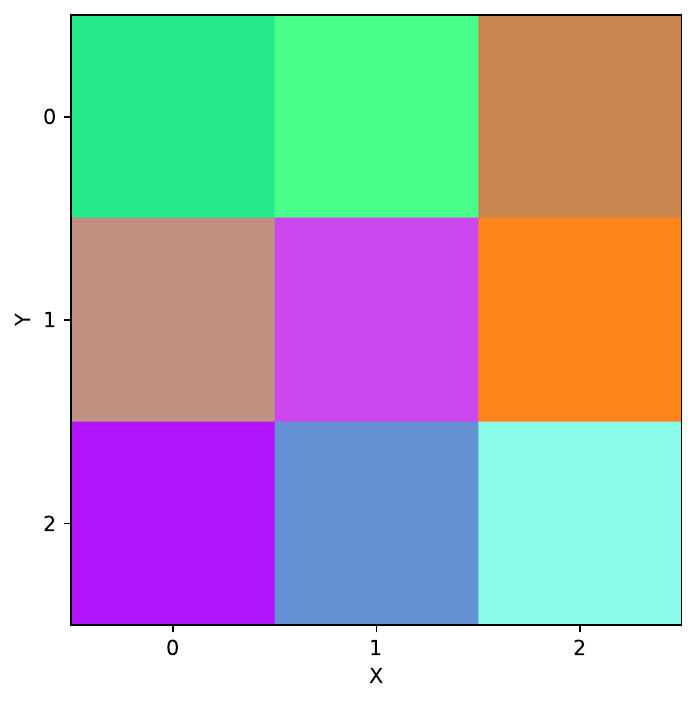}
         \caption{RGB}
         \label{fig:rgb}
     \end{subfigure}
     \caption{Grayscale and RGB images considered for example in the paper.}
\end{figure}

\begin{table}[!t]
\scriptsize
\centering
\caption{{The pixel values for the RGB image in Fig. \ref{fig:rgb}. The R values taken alone make the pixel values for the grayscale image in Fig. \ref{fig:gray}.}}
\label{tab:pixel_values}
\begin{tabular}{@{}ccccccccccc@{}}
\toprule
                            &            & \multicolumn{3}{c}{\textit{\textbf{R}}} & \multicolumn{3}{c}{\textit{\textbf{G}}} & \multicolumn{3}{c}{\textit{\textbf{B}}} \\ \midrule
\multirow{3}{*}{\textbf{Y}} & \textit{0} & 37          & 235         & 140         & 192         & 144         & 129         & 178         & 20          & 254         \\
                            & \textit{1} & 72          & 255         & 137         & 204         & 71          & 237         & 101         & 146         & 212         \\
                            & \textit{2} & 203         & 133         & 79          & 252         & 134         & 25          & 139         & 252         & 234         \\ \midrule 
                            &            & \textit{0}  & \textit{1}  & \textit{2}  & \textit{0}  & \textit{1}  & \textit{2}  & \textit{0}  & \textit{1}  & \textit{2}  \\ \cmidrule(l){3-11}
                            &            & \multicolumn{9}{c}{\textbf{X}}                                                                                              \\ \bottomrule
\end{tabular}
\end{table}

\subsection{FQRI}
Inspired by the qubit FRQI method, the flexible qutrit representation of quantum images (FQRI) is proposed. The image information is encoded in a quantum state given by the formula:
\begin{equation}
\label{eq:fqri}
    \left|I(\theta)\right>=\frac{1}{3^n}\sum_{i=0}^{3^{2n}-1}\left(\cos{\theta_i}\left|0\right>+\sin{\theta_i}\left|1\right>+0\left|2\right>\right)\otimes\left|i\right>
\end{equation}
$2n+1$ qutrits are required to encode a $3^n\times3^n$ image with $2n$ qutrits used to encode the location information, and the remaining qutrit is used to encode the pixel values. The pixel values are encoded in the $\theta_i$ values where $\theta_i\in[0,\pi/2]$. For an 8-bit grayscale image, the pixel values are in the range $[0,255]$, which are then scaled to be in $\theta$'s range.

The state can be prepared in two steps. First, the qutrit Hadamard gate $(H)$ is applied on the $2n$ qutrits to create a superposition state. Then $R_y^{(01)}(2\theta_i)$ gates controlled on the $2n$ location qutrits are applied. Formally:
\begin{enumerate}
    \item Begin with ${\left|0\right>}^{\otimes (2n+1)}$ state.
    \item Apply transform $\mathcal{H}=I\otimes H^{\otimes 2n}$:\\
    \begin{equation}
        \mathcal{H}{\left|0\right>}^{\otimes (2n+1)}=\frac{1}{3^n}\left|0\right>\otimes\sum_{i=0}^{3^{2n}-1}\left|i\right>=\left|H\right>
    \end{equation}
    \item Apply $2n$ controlled $R_y^{(01)}(2\theta_i)$ gates $(\mathcal{R}_i)$ for each pixel state $\ket{i}$:
    \begin{equation}
         \mathcal{R}_i=\left(I\otimes\sum_{j=0,j\ne i}^{3^{2n}-1}\left|j\right>\left<j\right|\right)+R_y^{(01)}(2\theta_i)\otimes\left|i\right>\left<i\right|
    \end{equation}
    \begin{equation}
        \left(\prod_{i=0}^{3^{2n}-1}\mathcal{R}_i\right)\left|H\right>=\left|I(\theta)\right>
    \end{equation}
\end{enumerate}

The circuit diagram pertinent to FQRI representation of the image in Fig. \ref{fig:gray} is provided in Fig. \ref{fig:fqri}.

The decoding of the encoded image is probabilistic and depends on the number of shots used. The final state can be measured in the computational basis to get the probabilities of all different basis states. The $\theta_i$ values can be calculated from the probabilities.

This encoding works well for grayscale images. As only two of the three possible states of the pixel value qutrit are being used, this encoding can be hypothesised to be implemented with a qubit-qutrit hybrid model where $2n$ qutrits are used for encoding the pixel location and a qubit is used for pixel values.

As mentioned earlier, all three states of the pixel value qutrit can be utilised to encode RGB information of colour images. This method is described in the following encoding.

\begin{figure}[!t]
\centering
    \begin{subfigure}{0.08\textheight}
         \centering
         \includegraphics[width=\textwidth]{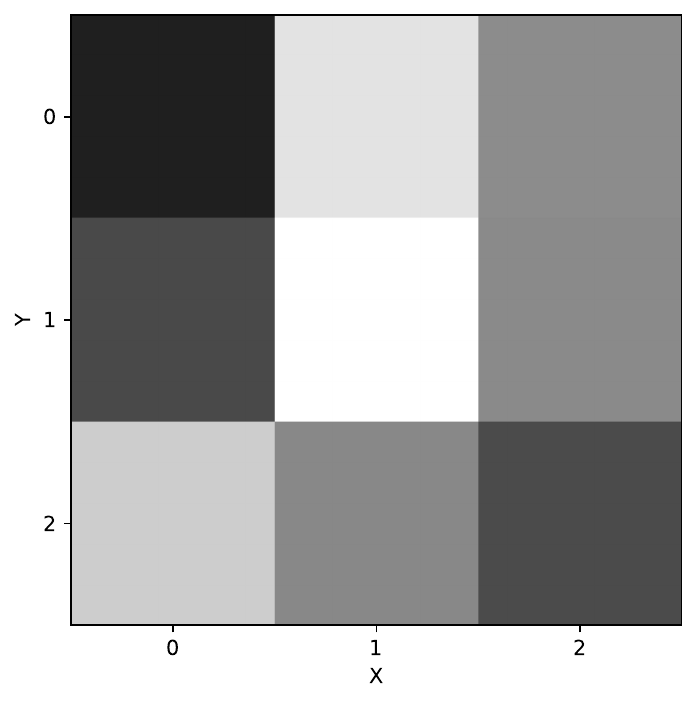}
         \caption{$10^4$}
    \end{subfigure}
    \begin{subfigure}{0.08\textheight}
         \centering
         \includegraphics[width=\textwidth]{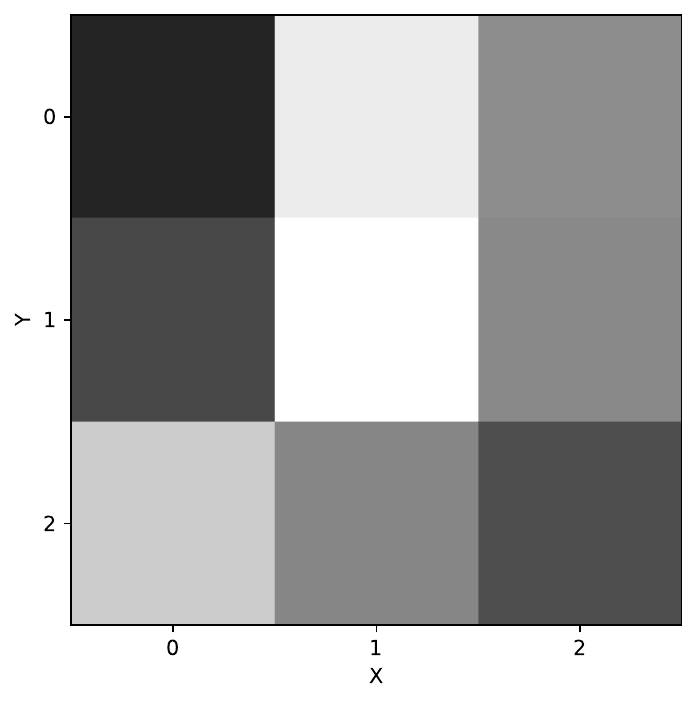}
         \caption{$10^5$}
    \end{subfigure}
    \begin{subfigure}{0.08\textheight}
         \centering
         \includegraphics[width=\textwidth]{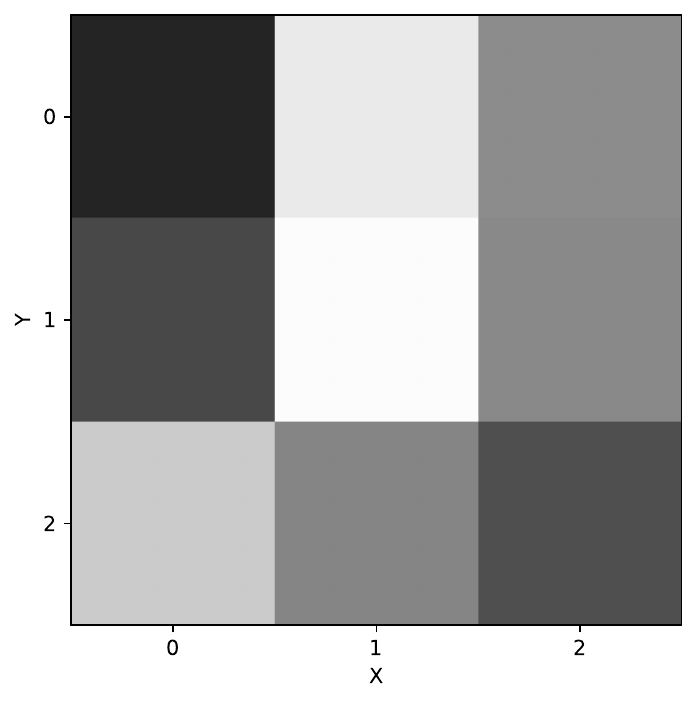}
         \caption{$10^6$}
     \end{subfigure}
     \caption{Decoded images using FQRI using different number of shots.}
     \label{fig:frqi}
\end{figure}

\subsection{FQRRI}
The flexible qutrit representation of RGB quantum images (FQRRI) is proposed here. In this encoding, $2n+1$ qutrits are used to encode RGB images. Again assuming 8-bit images, all R, G and B pixel values will be in the range $[0,255]$. The image is encoded in the quantum state given by:
\begin{equation}
\begin{split}
        \left|I(\theta)\right>=\frac{1}{3^n}\sum_{i=0}^{3^{2n}-1}\biggl(\cos{\theta_i^{gb}}\cos{\theta_i^{gr}}\left|0\right>+\sin{\theta_i^{gb}}\left|1\right>\biggr.\\
        +\cos{\theta_i^{gb}}\sin{\theta_i^{gr}}\left|2\right>\biggl.\biggr)\otimes\left|i\right>
\end{split}
\label{eq:qfrqi_c}
\end{equation}
Here, $\theta_i^{gb}$ and $\theta_i^{gr}$ encode the pixel values of G, B and G, R channels, respectively, of the $i^{th}$ pixel. The three values are encoded in two angles using the equations:
\begin{align}
    \theta_i^{gb}=\frac{(G_i\bmod16) \times 256+B_i}{4095}\times\frac{\pi}{2}\\
    \theta_i^{gr}=\frac{(G_i// 16) \times 256+R_i}{4095}\times\frac{\pi}{2}
\end{align}
The $//$ symbol represents the floor division operation. Because of the particular structure of these equations, all three pixel values for all pixels can be obtained.

The decoding is probabilistic and depends on the number of shots used. The state is measured in the computational basis to get counts of all basis states. Each measurement will provide us with a ternary string. From this string, the last $2n$ values give us the pixel location, and from the first value, the probability of getting $\left|0\right>,\left|1\right>$ and $\left|2\right>$ states can be calculated. Let's call these probabilities $p_0^i, p_1^i$ and $p_2^i$ for the $i^{th}$ pixel, then,
\begin{align}
    \theta_i^{gb}&=\sin^{-1}\left({3^n\times\sqrt{p_1^i}}\right)\\
    \theta_i^{gr}&=\tan^{-1}{\sqrt{\frac{p_2^i}{p_0^i}}}
\end{align}
This gives,
\begin{align}
    B_i&= \left(\theta_i^{gb}\times4095\times\frac{2}{\pi}\right)\bmod256\\
    R_i&= \left(\theta_i^{gr}\times4095\times\frac{2}{\pi}\right)\bmod256
\end{align}
\begin{equation}
    \begin{split}
           G_i&= \left[\left(\theta_i^{gb}\times4095\times\frac{2}{\pi}\right)//256\right] \\
    &+ \left[\left(\theta_i^{gr}\times4095\times\frac{2}{\pi}\right)\bmod256\right]\times16
    \end{split}
\end{equation}
To create the state in (\ref{eq:qfrqi_c}), the steps to be followed are:
\begin{enumerate}
    \item Begin with ${\left|0\right>}^{\otimes (2n+1)}$ state.
    \item Apply transform $\mathcal{H}=I\otimes H^{\otimes 2n}$:\\
    \begin{equation}
        \mathcal{H}{\left|0\right>}^{\otimes (2n+1)}=\frac{1}{3^n}\left|0\right>\otimes\sum_{i=0}^{3^{2n}-1}\left|i\right>=\left|H\right>
    \end{equation}
    \item Apply $2n$ controlled $R_y^{(01)}(2\theta_i^{gb})$ gates $(\mathcal{R}_i^{gb})$ and $2n$ controlled $R_y^{(02)}(2\theta_i^{gr})$ gates $(\mathcal{R}_i^{gr})$ for each pixel state $\ket{i}$:
    \begin{align}
         \mathcal{R}_i^{gb}&=\left(I\otimes\sum_{j=0,j\ne i}^{3^{2n}-1}\left|j\right>\left<j\right|\right)+R_y^{(01)}(2\theta_i^{gb})\otimes\left|i\right>\left<i\right|\\
         \mathcal{R}_i^{gr}&=\left(I\otimes\sum_{j=0,j\ne i}^{3^{2n}-1}\left|j\right>\left<j\right|\right)+R_y^{(02)}(2\theta_i^{gr})\otimes\left|i\right>\left<i\right|
    \end{align}
        \begin{equation}
        \left(\prod_{i=0}^{3^{2n}-1}\mathcal{R}_i^{gr}\mathcal{R}_i^{gb}\right)\left|H\right>=\left|I(\theta)\right>
    \end{equation}
\end{enumerate}

The circuit diagram for FQRRI of the image in Fig. \ref{fig:rgb} is shown in Fig. \ref{fig:fqrri}.

\begin{figure}[!t]
\centering
    \begin{subfigure}{0.08\textheight}
         \centering
         \includegraphics[width=\textwidth]{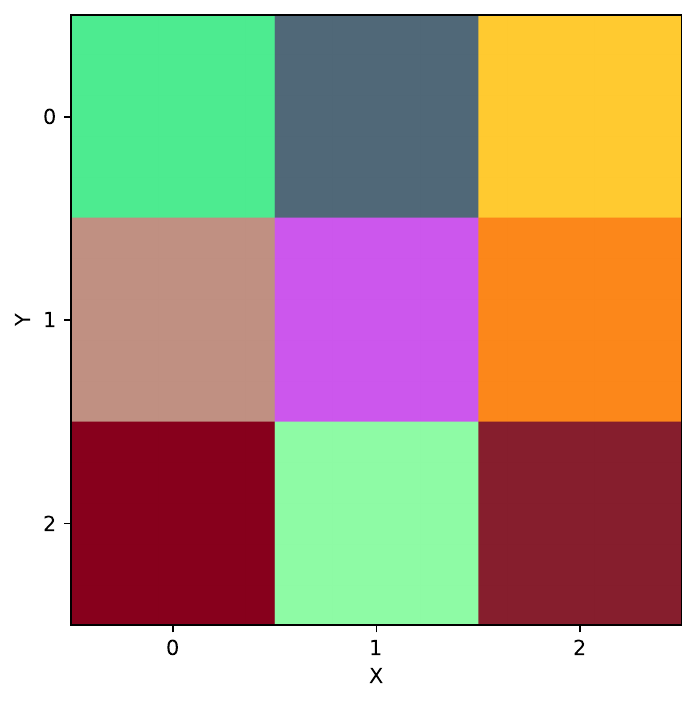}
         \caption{$10^4$}
    \end{subfigure}
    \begin{subfigure}{0.08\textheight}
         \centering
         \includegraphics[width=\textwidth]{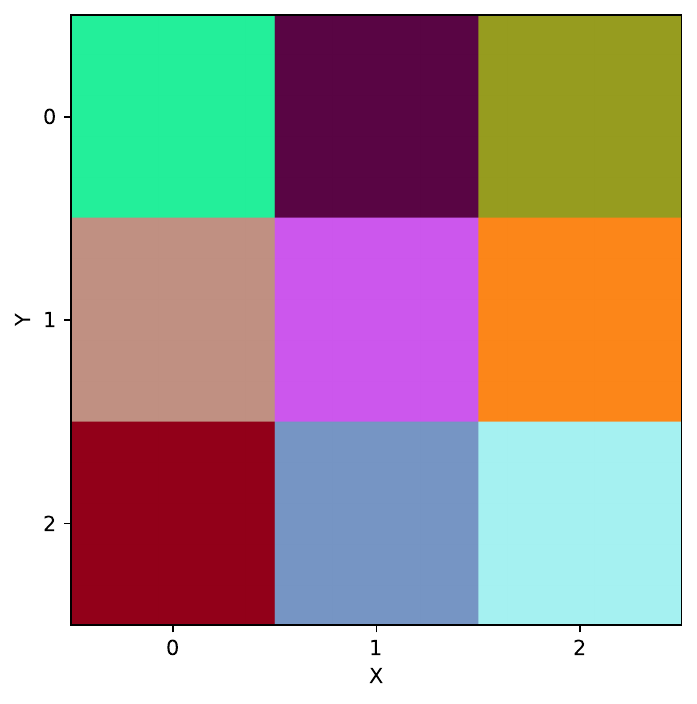}
         \caption{$10^5$}
    \end{subfigure}
    \begin{subfigure}{0.08\textheight}
         \centering
         \includegraphics[width=\textwidth]{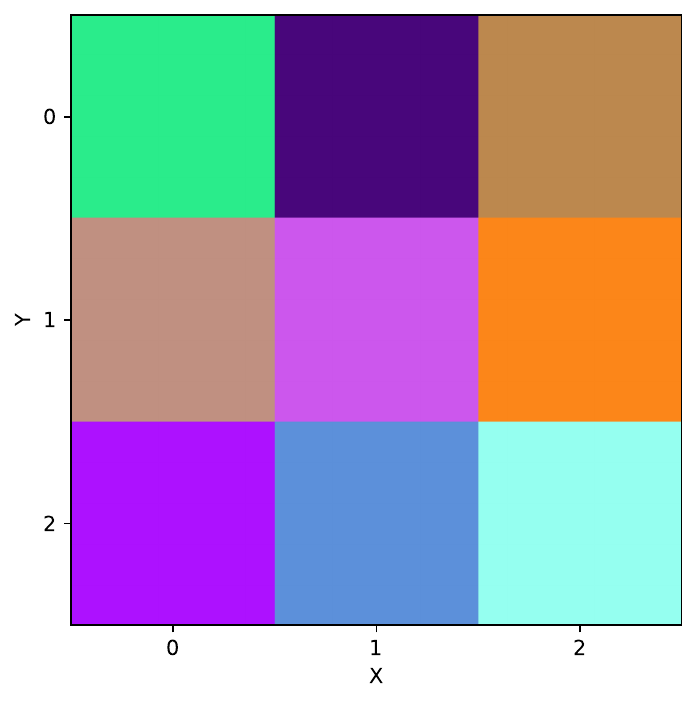}
         \caption{$10^6$}
     \end{subfigure}
     \caption{Decoded images using FQRRI using different number of shots.}
     \label{fig:frqi_c}
\end{figure}

\subsection{FQRQCI}
The qubit FRQCI method uses the phase to encode an extra value in the quantum state. This still leaves two angles to encode three values. This, like FQRRI, needs more shots to decode the image accurately. The extra state available in qutrits can be used to encode the three pixel values in three different angles. This inspires the flexible qutrit representation of quantum color images (FQRQCI).

As before, $2n+1$ qutrits are required to encode an RGB image. The quantum state of the image using this encoding is given by the formula:
\begin{equation}
\begin{split}
        \left|I(\theta)\right>=\frac{1}{3^n}\sum_{j=0}^{3^{2n}-1}\biggl(\cos{\theta_j^{r}}\left|0\right>+\sin{\theta_j^{r}}\cos{\theta_j^{g}}\left|1\right>\biggr.\\
        +e^{i\theta_j^b}\sin{\theta_j^{r}}\sin{\theta_j^{g}}\left|2\right>\biggl.\biggr)\otimes\left|j\right>
\end{split}
\label{eq:qfrqci}
\end{equation}
where,
\begin{align}
    \theta_j^r=\frac{R_j}{255}\times\frac{\pi}{2}\\
    \theta_j^g=\frac{G_j}{255}\times\frac{\pi}{2}\\
    \theta_j^b=\frac{B_j}{255}\times\frac{\pi}{2}
\end{align}
To create the state in (\ref{eq:qfrqci}), the steps to be followed are:
\begin{enumerate}
    \item Begin with ${\left|0\right>}^{\otimes (2n+1)}$ state.
    \item Apply transform $\mathcal{H}=I\otimes H^{\otimes 2n}$:\\
    \begin{equation}
        \mathcal{H}{\left|0\right>}^{\otimes (2n+1)}=\frac{1}{3^n}\left|0\right>\otimes\sum_{j=0}^{3^{2n}-1}\left|j\right>=\left|H\right>
    \end{equation}
    \item Apply $2n$ controlled $R_y^{(01)}(2\theta_j^{r})$ gates $(\mathcal{R}_j^{r})$ and $2n$ controlled $U^{(12)}(2\theta_j^{g},\theta_j^{b},0)$ gates $(\mathcal{U}_j^{gb})$ for each pixel state $\ket{j}$:
    \begin{equation}
         \mathcal{R}_j^{r}=\left(I\otimes\sum_{k=0,k\ne j}^{3^{2n}-1}\left|k\right>\left<k\right|\right)+R_y^{(01)}(2\theta_j^{r})\otimes\left|j\right>\left<j\right|
    \end{equation}
    \begin{equation}
         \mathcal{U}_j^{gb}=\left(I\otimes\sum_{k=0,k\ne j}^{3^{2n}-1}\left|k\right>\left<k\right|\right)+U^{(12)}(2\theta_j^{g},\theta_j^b,0)\otimes\left|j\right>\left<j\right|
    \end{equation}
    \begin{equation}
        \left(\prod_{j=0}^{3^{2n}-1}\mathcal{U}_j^{gb}\mathcal{R}_j^{r}\right)\left|H\right>=\left|I(\theta)\right>
    \end{equation}
\end{enumerate}

\begin{figure}[!t]
\centering
    \begin{subfigure}{0.08\textheight}
         \centering
         \includegraphics[width=\textwidth]{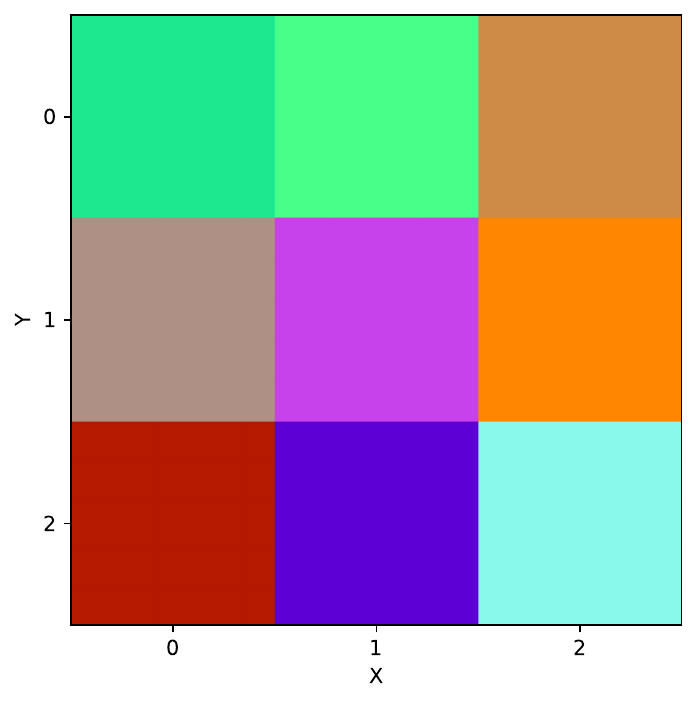}
         \caption{$10^4$}
    \end{subfigure}
    \begin{subfigure}{0.08\textheight}
         \centering
         \includegraphics[width=\textwidth]{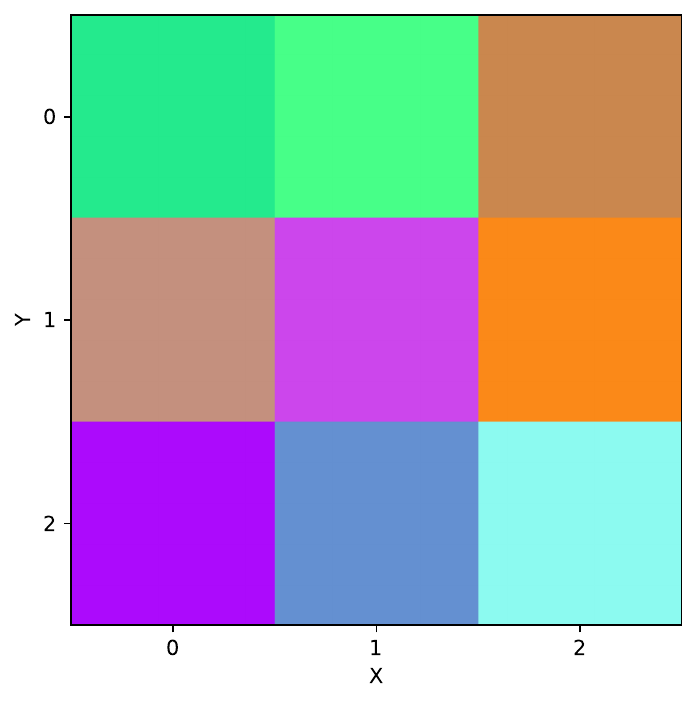}
         \caption{$10^5$}
    \end{subfigure}
    \begin{subfigure}{0.08\textheight}
         \centering
         \includegraphics[width=\textwidth]{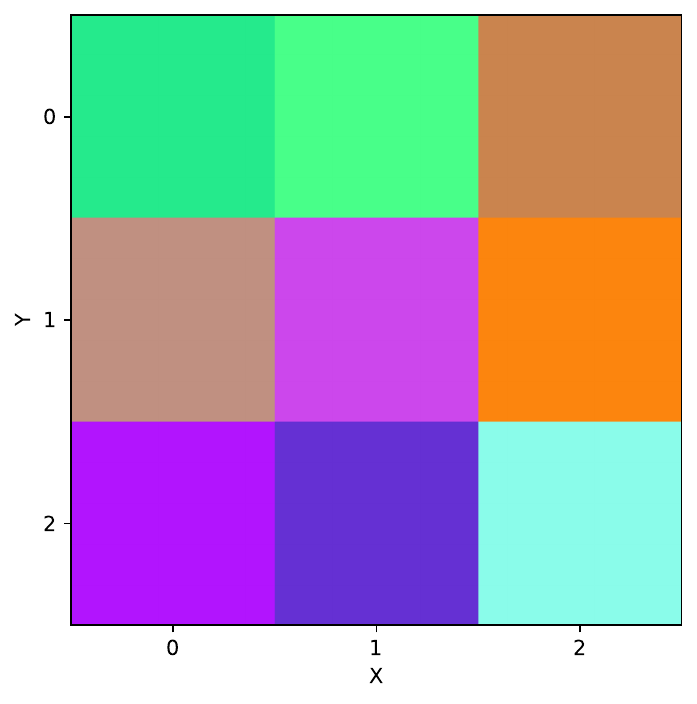}
         \caption{$10^6$}
     \end{subfigure}
     \caption{Decoded images using FQRQCI using different number of shots.}
     \label{fig:frqci}
\end{figure}

The circuit diagram for FQRQCI of the image in Fig. \ref{fig:rgb} is shown in Fig. \ref{fig:fqrqci}.

The decoding is probabilistic. To get back the pixel values, the following steps need to be performed:
\begin{enumerate}
    \item Measure the final state in the computational basis to get probabilities of $\left|0\right>$ and $\left|1\right>$ $(^1p_0^j,^1p_1^j)$ state of the pixel value qutrit, while the other qutrits provide information about the pixel location. Now,
  \begin{align}
      \theta^r_j&=\cos^{-1}\left({3^n\times\sqrt{^1p_0^j}}\right)\\
      \theta^g_j&=\cos^{-1}{\left(3^n\sqrt{\frac{^1p_1^j}{\sin^2{\theta^r_j}}}\right)}
  \end{align}
    \item Apply the operation $U^{(02)}(\frac{\pi}{2},-\pi,-\pi)$ on the pixel value qutrit and again measure the final state in the computational basis to get $(^2p_0^j,^2p_1^j)$.
    \item Separately, apply the operation $U^{(02)}(\frac{\pi}{2},-\frac{\pi}{2},\frac{\pi}{2})$ on the pixel value qutrit and again measure the final state in the computational basis to get $(^3p_0^j,^3p_1^j)$.
    \item Get the $\theta^b_j$ value as:
    \begin{equation}
        \theta^b_j=\tan^{-1}{\left(\frac{^3p_0^j-^3p_1^j}{^2p_0^j-^2p_1^j}\right)}
    \end{equation}
\end{enumerate}

\subsection{MCQRI}
The qubit MCQI method uses $2n+3$ qubits to encode colour images with $2n$ qubits used to encode pixel location, two qubits used to specify which colour to encode and the last qubit to encode the pixel value. The extra third state in qutrits can be used to specify which colour to use in a single qutrit. Thus, only $2n+2$ qutrits can be used to encode RGB images. Inspired by this, the multi channel qutrit representation for quantum images (MCQRI) is proposed.

The quantum state of the image is given by the formula:
\begin{equation}
\label{eq:qmcqi}
\begin{split}
    \left|I(\theta)\right>=\frac{1}{3^n\sqrt{3}}\sum_{i=0}^{3^{2n}-1}\biggl(\cos{\theta^i_r}\left|00\right>+\cos{\theta^i_g}\left|01\right>+\cos{\theta^i_b}\left|02\right>\\
    \sin{\theta^i_r}\left|10\right>+\sin{\theta^i_g}\left|11\right>+\sin{\theta^i_b}\left|12\right>\biggr)\otimes\left|i\right>
\end{split}
\end{equation}
The pixel values are encoded in the $\theta_j^i$ values where $\theta_j^i\in[0,\pi/2]$ by scaling. The following steps are followed to get to this state:
\begin{enumerate}
    \item Begin with ${\left|0\right>}^{\otimes (2n+2)}$ state.
    \item Apply transform $\mathcal{H}=I\otimes H^{\otimes 2n+1}$:\\
    \begin{equation}
        \mathcal{H}{\left|0\right>}^{\otimes (2n+2)}=\frac{1}{3^n\sqrt{3}}\left|0\right>\otimes\sum_{j=0}^{3^{2n}-1}\left|j\right>=\left|H\right>
    \end{equation}
    \item Apply three controlled $R_y^{(01)}(2\theta_j^{i})$ gates $(\mathcal{R}_j^{i})$ for each pixel state $\ket{i}$:
        \begin{align}
         \mathcal{R}_r^{i}&=\left(I\otimes\sum_{j=0,j\ne0}^{2}\left|j\right>\left<j\right|\right)+R_y^{(01)}(2\theta_r^{i})\otimes\left|0\right>\left<0\right|\\
         \mathcal{R}_g^{i}&=\left(I\otimes\sum_{j=0,j\ne1}^{2}\left|j\right>\left<j\right|\right)+R_y^{(01)}(2\theta_g^{i})\otimes\left|1\right>\left<1\right|\\
         \mathcal{R}_b^{i}&=\left(I\otimes\sum_{j=0,j\ne2}^{2}\left|j\right>\left<j\right|\right)+R_y^{(01)}(2\theta_b^{i})\otimes\left|2\right>\left<2\right|\\
         \mathcal{R}_i&=\mathcal{R}_b^{i}\mathcal{R}_g^{i}\mathcal{R}_r^{i}\\
         \mathbbm{R}_i&=\left(I^{\otimes2} \otimes\sum_{j=0,j\ne i}^{3^{2n}-1}\left|j\right>\left<j\right|\right)+\mathcal{R}_i\otimes\left|i\right>\left<i\right|
    \end{align}
    \begin{equation}
        \left(\prod_{i=0}^{3^{2n}-1}\mathbbm{R}_i\right)\left|H\right>=\left|I(\theta)\right>
    \end{equation}
    The $\mathcal{R}_j^i$ gates are controlled $R_y^{(01)}(2\theta_j^i)$ gates that are applied on the first two qutrits, which are in turn controlled by the last $2n$ qutrits.
\end{enumerate}

\begin{figure}[!t]
\centering
    \begin{subfigure}{0.08\textheight}
         \centering
         \includegraphics[width=\textwidth]{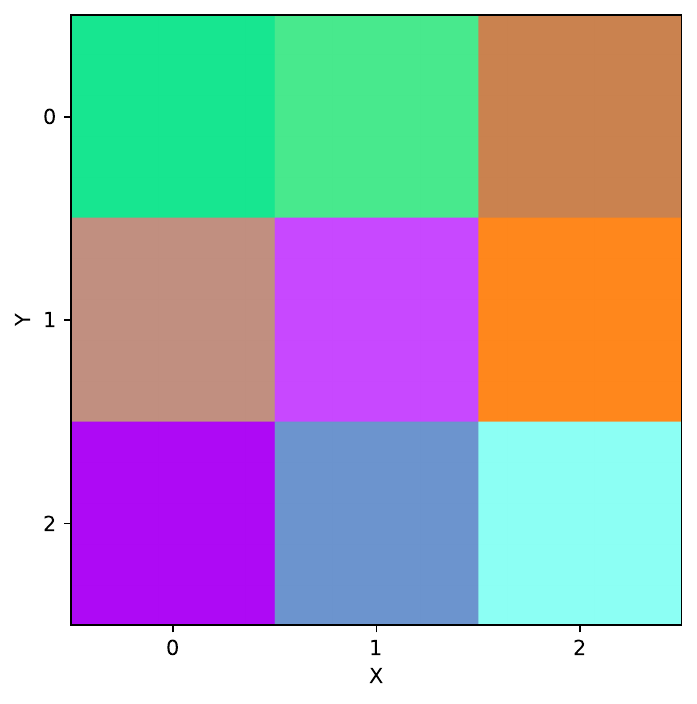}
         \caption{$10^4$}
    \end{subfigure}
    \begin{subfigure}{0.08\textheight}
         \centering
         \includegraphics[width=\textwidth]{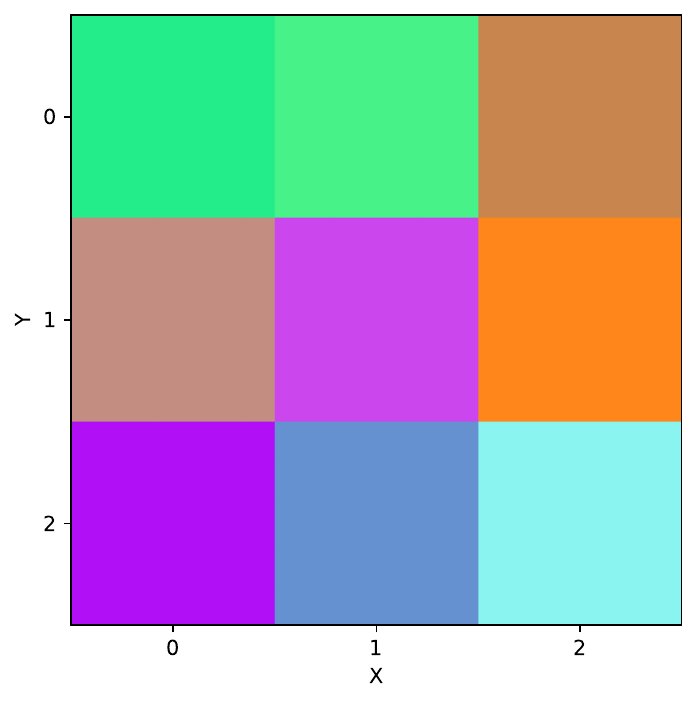}
         \caption{$10^5$}
    \end{subfigure}
    \begin{subfigure}{0.08\textheight}
         \centering
         \includegraphics[width=\textwidth]{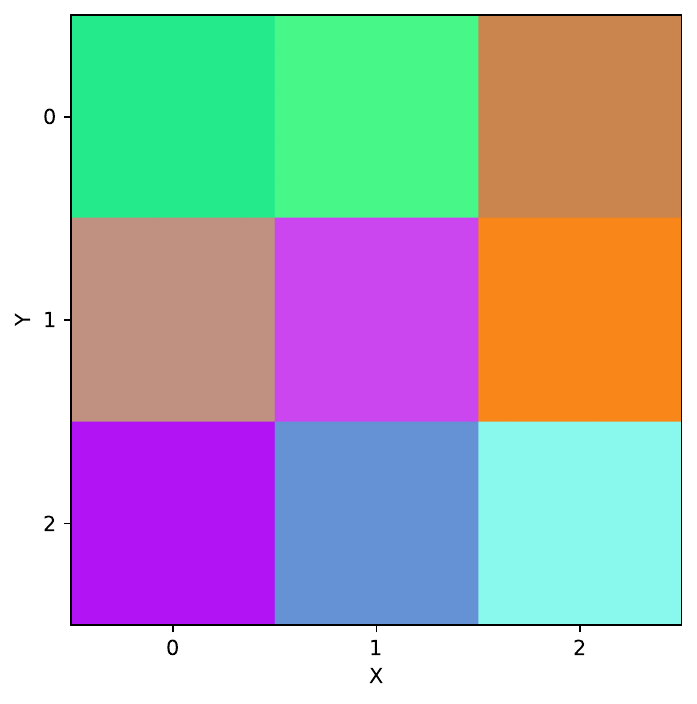}
         \caption{$10^6$}
     \end{subfigure}
     \caption{Decoded images using MCQRI using different number of shots.}
     \label{fig:mcqi}
\end{figure}

The circuit diagram for MCQRI of the image in Fig. \ref{fig:rgb} is shown in Fig. \ref{fig:mcqri}.

The decoding is probabilistic and depends on the number of shots used. The original image can be obtained by measuring the state in the computational basis. Let $\bm{p}$ denote the array of probabilities obtained after measuring the state. Assuming that the array is ordered in ascending order of the states, the size of the array will be $3^{2n+2}$. It can be divided into three equal parts. The first $3^{2n+1}$ values correspond to $\cos$ values and the next $3^{2n+1}$ values correspond to $\sin$ values. Let these be denoted by $\bm{p_1}$ and $\bm{p_2}$ respectively. The last $3^{2n+1}$ values will be zero. Now,
\begin{align}
    \bm{c^i} = \frac{1}{2}\cos^{-1}{\left(3^{2n+1}\times(\bm{p_1^i}-\bm{p_2^i})\right)}
\end{align}
These $\bm{c^i}$ values are the $\theta^i_j$ values from (\ref{eq:qmcqi}). Again, this array can be divided into three equal parts. The first part corresponds to R values, next to G values and the last to B values which can then be scaled back to the range $[0,255]$ to obtain the original image.

One can also think of implementing this as a qubit-qutrit hybrid model with $2n+1$ qutrits and one qubit.

\subsection{QRCIQ}
All the previous encodings had probabilistic decoding. The encoding considered here provides a deterministic decoding process. The QTRQ method is the qutrit extension of the qubit NEQR method. The quantum representation model of color digital images using qutrits (QRCIQ), the qutrit extension of the qubit QRCI method, is proposed.

The encoding requires $2n+2+3$ qutrits to encode RGB images. Here, $2n$ qutrits are required to encode pixel location, the next two qutrits are used to encode the bit plane number, and the remaining three qutrits encode the bit plane value. Assuming 8-bit images, the quantum state is given by the formula:
\begin{equation}
\label{eq:qrciq}
    \begin{split}
        \left|I\right>=\frac{1}{3^{n+1}}\sum_{b=0}^{5}\sum_{i=0}^{3^{2n}-1}\left|R_{b}^iG_{b}^iB_{b}^i\right>\otimes\left|i\right>
    \end{split}
\end{equation}
Here, $b$ represents the bit plane number and $R_{b}^iG_{b}^iB_{b}^i$ is the RGB bit value of the $i^{th}$ pixel of $b^{th}$ plane. The bit plane values are $\in\{0,1,2\}$. The state is prepared using the following steps:
\begin{enumerate}
    \item Begin with ${\left|0\right>}^{\otimes (2n+2+3)}$ state.
    \item Apply transform $\mathcal{H}=I\otimes I\otimes I\otimes H^{\otimes 2n+2}$:\\
    \begin{equation}
        \mathcal{H}{\left|0\right>}^{\otimes (2n+2)}=\frac{1}{3^{n+1}}{\left|000\right>}\otimes\sum_{j=0}^{3^{2n}-1}\left|j\right>=\left|H\right>
    \end{equation}
    \item If $C_b^i=0$, where $C\in\{R,G,B\}$, do nothing.
    \item If $C_b^i=1$, apply $\left[+1\right]$ gate on the $C$ qutrit controlled by $2n+2$ qutrits with control state $\ket{b}\otimes\ket{i}$.
    \item If $C_b^i=2$, apply $\left[+2\right]$ gate on the $C$ qutrit controlled by $2n+2$ qutrits with control state $\ket{b}\otimes\ket{i}$.
\end{enumerate}

The circuit diagram for QRCIQ of the image in Fig. \ref{fig:rgb} is shown in Fig. \ref{fig:qrciq}.

The decoding is deterministic as after measuring the state in the computational basis, the mere presence of a basis state provides all information it encodes. The bit string of the basis state can be decoded to give the pixel location, bit plane number and the bit plane pixel value. It is important to note here that extra states will be present in the final state corresponding to unused bit planes. As two qutrits can encode $3^2=9$ bit planes and only six are required for an 8-bit image using ternary strings, three bit planes are empty, corresponding to $b=6,7$ or $8$. Thus, while decoding, these states need to be discarded. It's possible to envision that these available "slots" can be populated to provide extra information about the images or facilitate processing or error correction.

\begin{figure}[!t]
\centering
    \begin{subfigure}{0.08\textheight}
         \centering
         \includegraphics[width=\textwidth]{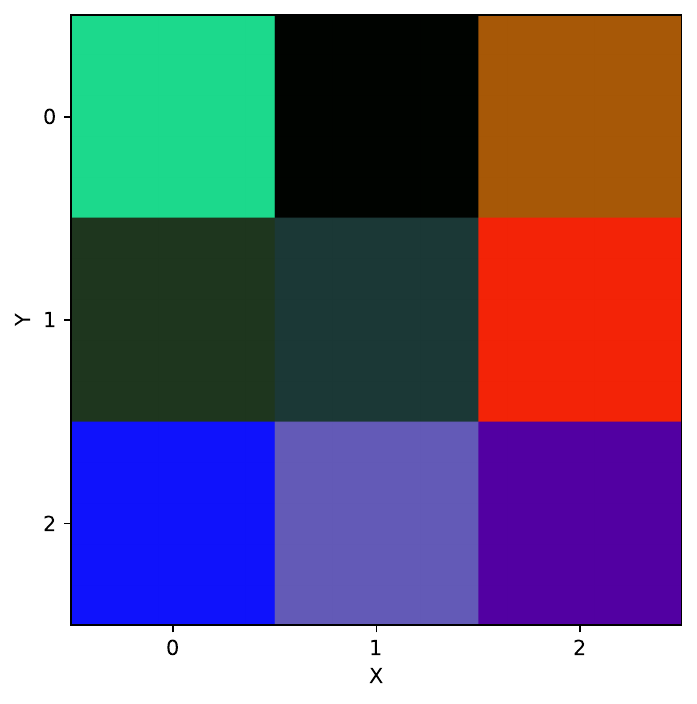}
         \caption{$50$}
    \end{subfigure}
    \begin{subfigure}{0.08\textheight}
         \centering
         \includegraphics[width=\textwidth]{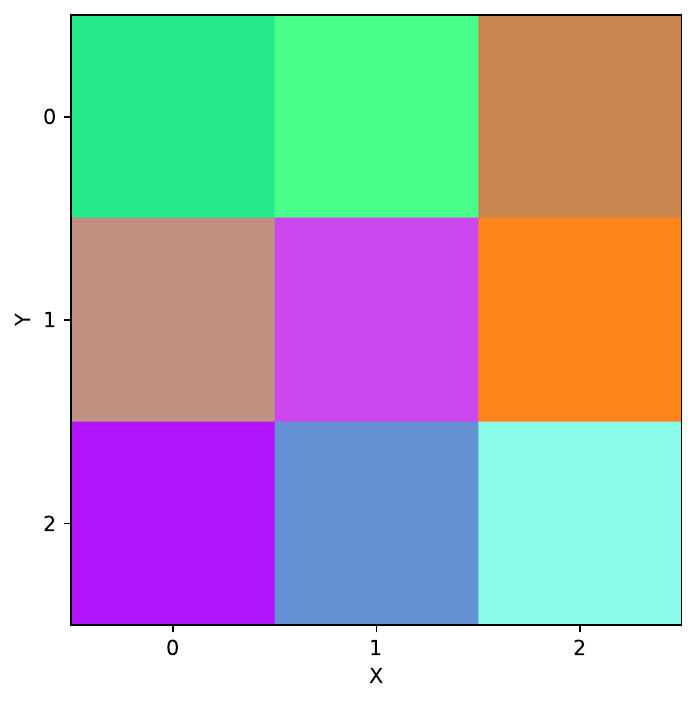}
         \caption{$500$}
    \end{subfigure}
    \caption{Decoded images using QRCIQ using different number of shots.}
    \label{fig:qrci}
\end{figure}

\subsection*{Remark}
The encodings that have a probabilistic decoding, use inverse trigonometric functions to calculate the encoded pixel values. These functions have a finite domain, for example, $sin^{-1}: [-1,1]\rightarrow[-\frac{\pi}{2},\frac{\pi}{2}]$. The values calculated using a finite number of shots may fall outside the domain of these functions. Thus, the values need to be clipped to stay in the domain to get a valid result. The accuracy of the calculated values can be improved by increasing the number of shots. While QRCIQ does not have a probabilistic decoding and does not depend on the number of shots used, it still requires a minimum number of shots to observe all basis states at least once. Once all states have been observed, increasing the number of shots will have no effect on the decoding.

\section{Conclusion and Future Work}
\label{conclude}
\begin{figure}[]
\centering
    \includegraphics[width=\columnwidth]{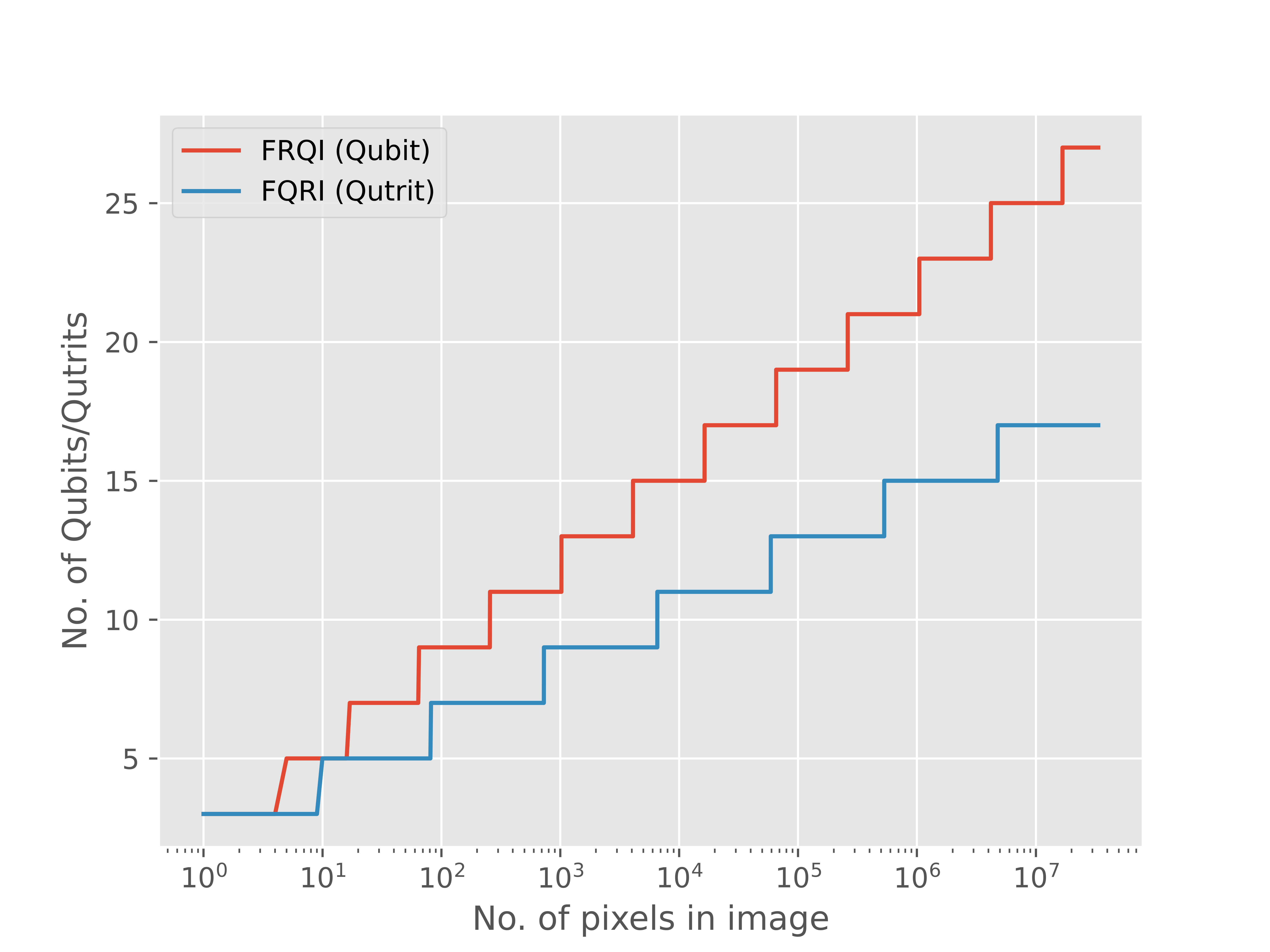}
    \caption{Number of qubits or qutrits required to represent an image.}
    \label{fig:compare}
\end{figure}

A few implementations to represent images as quantum states using qutrits have been described. All of these were validated using simulations with Cirq.
While the decoding is probabilistic for most of them and depends on the number of shots used, it is also important to note that decoding the image is not the only use case for these representations.
Different image processing tasks can be performed on the final quantum state \cite{https://doi.org/10.48550/arxiv.2203.01831,Ruan2021,Yan_Venegas-Andraca_2020,6379238}.
Machine learning tasks like classification can also be performed on the qutrit quantum representation of the images as performed with qubit QIRs \cite{khandelwal2022classifying, https://doi.org/10.48550/arxiv.2110.05476}.
A summary of the provided methods is shown in Table \ref{tab:summary}. The resource requirement increases slower than the qubit case and is always lower except for trivial cases. This can be seen from Fig. \ref{fig:compare}.

\begin{table}[b]
\centering
\caption{Comparison of the different quantum image representation methods presented. The number of required qutrits are for an image of size $3^n\times3^n$.}
\label{tab:summary}
\begin{tabular}{@{}ccccc@{}}
\toprule
\textbf{QIR}    & \textbf{\begin{tabular}[c]{@{}c@{}}Color \\ Space\end{tabular}} & \textbf{\begin{tabular}[c]{@{}c@{}}Required\\  Qutrits\end{tabular}} & \textbf{\begin{tabular}[c]{@{}c@{}}Decoding\\  Method\end{tabular}} & \textbf{Equation} \\ \midrule
\textit{FQRI}   & Grayscale            & $2n+1$                     & Probabilistic            & Eq. \ref{eq:fqri}              \\
\textit{FQRRI}  & RGB                  & $2n+1$                     & Probabilistic            & Eq. \ref{eq:qfrqi_c}           \\
\textit{FQRQCI} & RGB                  & $2n+1$                     & Probabilistic            & Eq. \ref{eq:qfrqci}            \\
\textit{MCQRI}  & RGB                  & $2n+2$                     & Probabilistic            & Eq. \ref{eq:qmcqi}             \\
\textit{QRCIQ}  & RGB                  & $2n+5$                     & Deterministic            & Eq. \ref{eq:qrciq}             \\ \bottomrule
\end{tabular}
\end{table}

While from the universality of qudit logic, the decomposition of the multi-qutrit gates used can be assured, the decompositions have not been looked at in this paper. Research can be done to find optimal decompositions of the gates. More representations can be studied to utilise the extra state available in qutrits. Appropriate use of the empty slots can also be studied. Comparison of different representations on their effectiveness in various image processing tasks can also be explored.

\bibliographystyle{IEEEtran}
\bibliography{IEEEabrv,refrences}

\iftrue
\begin{rotatepage}
\begin{sidewaysfigure*}[p]
    \centering
    \resizebox{.5\linewidth}{!}{
\Qcircuit @R=1em @C=0.75em {
 \\
 &\lstick{\ket{0}}& \qw&                \qw&\gate{{R_y^{(01)}(0.45)}} \qw    &\gate{{R_y^{(01)}(2.88)}} \qw    &\gate{{R_y^{(01)}(1.72)}} \qw    &\gate{{R_y^{(01)}(0.88)}} \qw    &\gate{{R_y^{(01)}(3.13)}} \qw    &\gate{{R_y^{(01)}(1.68)}} \qw    &\gate{{R_y^{(01)}(2.49)}} \qw    &\gate{{R_y^{(01)}(1.63)}} \qw    &\gate{{R_y^{(01)}(0.97)}} \qw        &\qwa\\
 &\lstick{\ket{0}}& \qw&\gate{{H}} \qw&  \push{\circled{0}}        \qw\qwx&  \push{\circled{0}}      \qw\qwx&  \push{\circled{0}}      \qw\qwx&  \push{\circled{1}}           \qw\qwx&  \push{\circled{1}}           \qw\qwx&  \push{\circled{1}}           \qw\qwx&  \push{\circled{2}}      \qw\qwx&  \push{\circled{2}}      \qw\qwx&  \push{\circled{2}}      \qw\qwx&       \qwa\\
 &\lstick{\ket{0}}& \qw&\gate{{H}} \qw&  \push{\circled{0}}      \qw\qwx&  \push{\circled{1}}           \qw\qwx&  \push{\circled{2}}      \qw\qwx&  \push{\circled{0}}      \qw\qwx&  \push{\circled{1}}           \qw\qwx&  \push{\circled{2}}      \qw\qwx&  \push{\circled{0}}      \qw\qwx&  \push{\circled{1}}           \qw\qwx&  \push{\circled{2}}      \qw\qwx&       \qwa\\
 \\
}}
    \caption{The circuit diagram for the FQRI representation for the image in Fig. \ref{fig:gray}.}
    \label{fig:fqri}
    \resizebox{\linewidth}{!}{\Qcircuit @R=1em @C=0.75em {
 \\
 &\lstick{\ket{0}}& \qw&                \qw&\gate{{R_y^{(01)}(0.14)}} \qw    &\gate{{R_y^{(02)}(2.39)}} \qw    &\gate{{R_y^{(01)}(0.02)}} \qw    &\gate{{R_y^{(02)}(1.95)}} \qw    &\gate{{R_y^{(01)}(0.39)}} \qw    &\gate{{R_y^{(02)}(1.68)}} \qw    &\gate{{R_y^{(01)}(2.43)}} \qw    &\gate{{R_y^{(02)}(2.41)}} \qw    &\gate{{R_y^{(01)}(1.49)}} \qw    &\gate{{R_y^{(02)}(0.98)}} \qw    &\gate{{R_y^{(01)}(2.72)}} \qw    &\gate{{R_y^{(02)}(2.85)}} \qw    &\gate{{R_y^{(01)}(2.46)}} \qw    &\gate{{R_y^{(02)}(3.1)}} \qw    &\gate{{R_y^{(01)}(1.37)}} \qw    &\gate{{R_y^{(02)}(1.67)}} \qw    &\gate{{R_y^{(01)}(1.95)}} \qw    &\gate{{R_y^{(02)}(0.26)}} \qw      &\qwa\\
 &\lstick{\ket{0}}& \qw&\gate{{H}} \qw&\push{\circled{0}}      \qw\qwx&\push{\circled{0}}      \qw\qwx&\push{\circled{0}}      \qw\qwx&\push{\circled{0}}      \qw\qwx&\push{\circled{0}}      \qw\qwx&\push{\circled{0}}      \qw\qwx&\push{\circled{1}}         \qw\qwx&\push{\circled{1}}         \qw\qwx&\push{\circled{1}}         \qw\qwx&\push{\circled{1}}         \qw\qwx&\push{\circled{1}}         \qw\qwx&\push{\circled{1}}         \qw\qwx&\push{\circled{2}}      \qw\qwx&\push{\circled{2}}     \qw\qwx&\push{\circled{2}}      \qw\qwx&\push{\circled{2}}      \qw\qwx&\push{\circled{2}}      \qw\qwx&\push{\circled{2}}      \qw\qwx&       \qwa\\
 &\lstick{\ket{0}}& \qw&\gate{{H}} \qw&\push{\circled{0}}      \qw\qwx&\push{\circled{0}}      \qw\qwx&\push{\circled{1}}         \qw\qwx&\push{\circled{1}}         \qw\qwx&\push{\circled{2}}      \qw\qwx&\push{\circled{2}}      \qw\qwx&\push{\circled{0}}      \qw\qwx&\push{\circled{0}}      \qw\qwx&\push{\circled{1}}         \qw\qwx&\push{\circled{1}}         \qw\qwx&\push{\circled{2}}      \qw\qwx&\push{\circled{2}}      \qw\qwx&\push{\circled{0}}      \qw\qwx&\push{\circled{0}}     \qw\qwx&\push{\circled{1}}         \qw\qwx&\push{\circled{1}}         \qw\qwx&\push{\circled{2}}      \qw\qwx&\push{\circled{2}}      \qw\qwx&       \qwa\\
 \\
}}
    \caption{The circuit diagram for the FQRRI representation for the image in Fig. \ref{fig:rgb}.}
    \label{fig:fqrri}
    \resizebox{\linewidth}{!}{\Qcircuit @R=1em @C=0.75em {
 \\
 &\lstick{\ket{0}}& \qw&                \qw&\gate{{R_y^{(01)}(0.46)}} \qw    &\gate{{U^{(12)}(2.37, 2.19, 0)}} \qw    &\gate{{R_y^{(01)}(2.9)}} \qw    &\gate{{U^{(12)}(1.77, 0.25, 0)}} \qw    &\gate{{R_y^{(01)}(1.72)}} \qw    &\gate{{U^{(12)}(1.59, 3.13, 0)}} \qw    &\gate{{R_y^{(01)}(0.89)}} \qw    &\gate{{U^{(12)}(2.51, 1.24, 0)}} \qw    &\gate{{R_y^{(01)}(3.14)}} \qw    &\gate{{U^{(12)}(0.87, 1.8, 0)}} \qw    &\gate{{R_y^{(01)}(1.69)}} \qw    &\gate{{U^{(12)}(2.92, 2.61, 0)}} \qw    &\gate{{R_y^{(01)}(2.5)}} \qw    &\gate{{U^{(12)}(3.1, 1.71, 0)}} \qw    &\gate{{R_y^{(01)}(1.64)}} \qw    &\gate{{U^{(12)}(1.65, 3.1, 0)}} \qw    &\gate{{R_y^{(01)}(0.97)}} \qw    &\gate{{U^{(12)}(0.31, 2.88, 0)}} \qw    &\qwa\\
 &\lstick{\ket{0}}& \qw&\gate{{H}} \qw&\push{\circled{0}}        \qw\qwx&\push{\circled{0}}                  \qw\qwx&\push{\circled{0}}       \qw\qwx&\push{\circled{0}}                  \qw\qwx&\push{\circled{0}}        \qw\qwx&\push{\circled{0}}                  \qw\qwx&\push{\circled{1}}         \qw\qwx&\push{\circled{1}}                   \qw\qwx&\push{\circled{1}}         \qw\qwx&\push{\circled{1}}                  \qw\qwx&\push{\circled{1}}         \qw\qwx&\push{\circled{1}}                   \qw\qwx&\push{\circled{2}}       \qw\qwx&\push{\circled{2}}                 \qw\qwx&\push{\circled{2}}        \qw\qwx&\push{\circled{2}}                 \qw\qwx&\push{\circled{2}}        \qw\qwx&\push{\circled{2}}                  \qw\qwx&                             \qwa\\
 &\lstick{\ket{0}}& \qw&\gate{{H}} \qw&\push{\circled{0}}        \qw\qwx&\push{\circled{0}}                  \qw\qwx&\push{\circled{1}}        \qw\qwx&\push{\circled{1}}                   \qw\qwx&\push{\circled{2}}        \qw\qwx&\push{\circled{2}}                  \qw\qwx&\push{\circled{0}}        \qw\qwx&\push{\circled{0}}                  \qw\qwx&\push{\circled{1}}         \qw\qwx&\push{\circled{1}}                  \qw\qwx&\push{\circled{2}}        \qw\qwx&\push{\circled{2}}                  \qw\qwx&\push{\circled{0}}       \qw\qwx&\push{\circled{0}}                 \qw\qwx&\push{\circled{1}}         \qw\qwx&\push{\circled{1}}                  \qw\qwx&\push{\circled{2}}        \qw\qwx&\push{\circled{2}}                  \qw\qwx&                                \qwa\\
 \\
}}
    \caption{The circuit diagram for the FQRQCI representation for the image in Fig. \ref{fig:rgb}.}
    \label{fig:fqrqci}
    \resizebox{\linewidth}{!}{\Qcircuit @R=1em @C=0.75em {
 \\
 &\lstick{\ket{0}}& \qw&                \qw&\gate{{R_y^{(01)}(0.46)}} \qw    &\gate{{R_y^{(01)}(0.89)}} \qw    &\gate{{R_y^{(01)}(2.5)}} \qw    &\gate{{R_y^{(01)}(2.9)}} \qw    &\gate{{R_y^{(01)}(3.14)}} \qw    &\gate{{R_y^{(01)}(1.64)}} \qw    &\gate{{R_y^{(01)}(1.72)}} \qw    &\gate{{R_y^{(01)}(1.69)}} \qw    &\gate{{R_y^{(01)}(0.97)}} \qw    &\gate{{R_y^{(01)}(2.37)}} \qw    &\gate{{R_y^{(01)}(2.51)}} \qw    &\gate{{R_y^{(01)}(3.1)}} \qw    &\gate{{R_y^{(01)}(1.77)}} \qw    &\gate{{R_y^{(01)}(0.87)}} \qw    &\gate{{R_y^{(01)}(1.65)}} \qw    &\gate{{R_y^{(01)}(1.59)}} \qw    &\gate{{R_y^{(01)}(2.92)}} \qw    &\gate{{R_y^{(01)}(0.31)}} \qw    &\gate{{R_y^{(01)}(2.19)}} \qw    &\gate{{R_y^{(01)}(1.24)}} \qw    &\gate{{R_y^{(01)}(1.71)}} \qw    &\gate{{R_y^{(01)}(0.25)}} \qw    &\gate{{R_y^{(01)}(1.8)}} \qw    &\gate{{R_y^{(01)}(3.1)}} \qw    &\gate{{R_y^{(01)}(3.13)}} \qw    &\gate{{R_y^{(01)}(2.61)}} \qw    &\gate{{R_y^{(01)}(2.88)}} \qw      &\qwa\\
 &\lstick{\ket{0}}& \qw&\gate{{H}} \qw&\push{\circled{0}}       \qw\qwx&\push{\circled{1}}          \qw\qwx&\push{\circled{2}}      \qw\qwx&\push{\circled{0}}      \qw\qwx&\push{\circled{1}}          \qw\qwx&\push{\circled{2}}       \qw\qwx&\push{\circled{0}}       \qw\qwx&\push{\circled{1}}          \qw\qwx&\push{\circled{2}}       \qw\qwx&\push{\circled{0}}       \qw\qwx&\push{\circled{1}}          \qw\qwx&\push{\circled{2}}      \qw\qwx&\push{\circled{0}}       \qw\qwx&\push{\circled{1}}          \qw\qwx&\push{\circled{2}}       \qw\qwx&\push{\circled{0}}       \qw\qwx&\push{\circled{1}}          \qw\qwx&\push{\circled{2}}       \qw\qwx&\push{\circled{0}}       \qw\qwx&\push{\circled{1}}          \qw\qwx&\push{\circled{2}}       \qw\qwx&\push{\circled{0}}       \qw\qwx&\push{\circled{1}}         \qw\qwx&\push{\circled{2}}      \qw\qwx&\push{\circled{0}}       \qw\qwx&\push{\circled{1}}          \qw\qwx&\push{\circled{2}}       \qw\qwx&      \qwa\\
 &\lstick{\ket{0}}& \qw&\gate{{H}} \qw&\push{\circled{0}}       \qw\qwx&\push{\circled{0}}       \qw\qwx&\push{\circled{0}}      \qw\qwx&\push{\circled{1}}         \qw\qwx&\push{\circled{1}}          \qw\qwx&\push{\circled{1}}          \qw\qwx&\push{\circled{2}}       \qw\qwx&\push{\circled{2}}       \qw\qwx&\push{\circled{2}}       \qw\qwx&\push{\circled{0}}       \qw\qwx&\push{\circled{0}}       \qw\qwx&\push{\circled{0}}      \qw\qwx&\push{\circled{1}}          \qw\qwx&\push{\circled{1}}          \qw\qwx&\push{\circled{1}}          \qw\qwx&\push{\circled{2}}       \qw\qwx&\push{\circled{2}}       \qw\qwx&\push{\circled{2}}       \qw\qwx&\push{\circled{0}}       \qw\qwx&\push{\circled{0}}       \qw\qwx&\push{\circled{0}}       \qw\qwx&\push{\circled{1}}          \qw\qwx&\push{\circled{1}}         \qw\qwx&\push{\circled{1}}         \qw\qwx&\push{\circled{2}}       \qw\qwx&\push{\circled{2}}       \qw\qwx&\push{\circled{2}}       \qw\qwx&      \qwa\\
 &\lstick{\ket{0}}& \qw&\gate{{H}} \qw&\push{\circled{0}}       \qw\qwx&\push{\circled{0}}       \qw\qwx&\push{\circled{0}}      \qw\qwx&\push{\circled{0}}      \qw\qwx&\push{\circled{0}}       \qw\qwx&\push{\circled{0}}       \qw\qwx&\push{\circled{0}}       \qw\qwx&\push{\circled{0}}       \qw\qwx&\push{\circled{0}}       \qw\qwx&\push{\circled{1}}          \qw\qwx&\push{\circled{1}}          \qw\qwx&\push{\circled{1}}         \qw\qwx&\push{\circled{1}}          \qw\qwx&\push{\circled{1}}          \qw\qwx&\push{\circled{1}}          \qw\qwx&\push{\circled{1}}          \qw\qwx&\push{\circled{1}}          \qw\qwx&\push{\circled{1}}          \qw\qwx&\push{\circled{2}}       \qw\qwx&\push{\circled{2}}       \qw\qwx&\push{\circled{2}}       \qw\qwx&\push{\circled{2}}       \qw\qwx&\push{\circled{2}}      \qw\qwx&\push{\circled{2}}      \qw\qwx&\push{\circled{2}}       \qw\qwx&\push{\circled{2}}       \qw\qwx&\push{\circled{2}}       \qw\qwx& \qwa\\
 \\
}
}
    \caption{The circuit diagram for the MCQRI representation for the image in Fig. \ref{fig:rgb}.}
    \label{fig:mcqri}
    \resizebox{\linewidth}{!}{\Qcircuit @R=1em @C=0.75em {
 \\
 &\lstick{\ket{0}}& \qw&                \qw&                   \qw    &\gate{\text{+1}} \qw    &                   \qw    &                   \qw    &                   \qw    &                   \qw    &                   \qw    &\gate{\text{+2}} \qw    &                   \qw    &\gate{\text{+2}} \qw    &                   \qw    &                   \qw    &\gate{\text{+2}} \qw    &                   \qw    &                   \qw    &\gate{\text{+2}} \qw    &\gate{\text{+1}} \qw    &                   \qw    &                   \qw    &\gate{\text{+1}} \qw    &                   \qw    &\gate{\text{+1}} \qw    &                   \qw    &                   \qw    &\gate{\text{+2}} \qw    &                   \qw    &\gate{\text{+1}} \qw    &                   \qw    &                   \qw    &\gate{\text{+1}} \qw    &                   \qw    &                   \qw    &\gate{\text{+1}} \qw    &                   \qw    &                   \qw    &                   \qw    &                   \qw    &                   \qw    &\gate{\text{+2}} \qw    &                   \qw    &\gate{\text{+1}} \qw    &                   \qw    &\gate{\text{+2}} \qw    &                   \qw  &\rstick{\cdots} \qw  \\
 &\lstick{\ket{0}}& \qw&                \qw&\gate{\text{+1}} \qw    &                   \qw\qwx&                   \qw    &\gate{\text{+1}} \qw    &\gate{\text{+2}} \qw    &                   \qw    &                   \qw    &                   \qw\qwx&\gate{\text{+1}} \qw    &                   \qw\qwx&\gate{\text{+1}} \qw    &                   \qw    &                   \qw\qwx&                   \qw    &\gate{\text{+1}} \qw    &                   \qw\qwx&                   \qw\qwx&\gate{\text{+1}} \qw    &                   \qw    &                   \qw\qwx&                   \qw    &                   \qw\qwx&\gate{\text{+2}} \qw    &                   \qw    &                   \qw\qwx&                   \qw    &                   \qw\qwx&\gate{\text{+1}} \qw    &                   \qw    &                   \qw\qwx&\gate{\text{+2}} \qw    &                   \qw    &                   \qw\qwx&\gate{\text{+2}} \qw    &                   \qw    &\gate{\text{+1}} \qw    &\gate{\text{+2}} \qw    &                   \qw    &                   \qw\qwx&                   \qw    &                   \qw\qwx&\gate{\text{+2}} \qw    &                   \qw\qwx&\gate{\text{+1}} \qw  &\rstick{\cdots} \qw   \\
 &\lstick{\ket{0}}& \qw&                \qw&                   \qw\qwx&                   \qw\qwx&\gate{\text{+1}} \qw    &                   \qw\qwx&                   \qw\qwx&\gate{\text{+1}} \qw    &\gate{\text{+1}} \qw    &                   \qw\qwx&                   \qw\qwx&                   \qw\qwx&                   \qw\qwx&\gate{\text{+1}} \qw    &                   \qw\qwx&\gate{\text{+2}} \qw    &                   \qw\qwx&                   \qw\qwx&                   \qw\qwx&                   \qw\qwx&\gate{\text{+2}} \qw    &                   \qw\qwx&\gate{\text{+2}} \qw    &                   \qw\qwx&                   \qw\qwx&\gate{\text{+2}} \qw    &                   \qw\qwx&\gate{\text{+2}} \qw    &                   \qw\qwx&                   \qw\qwx&\gate{\text{+2}} \qw    &                   \qw\qwx&                   \qw\qwx&\gate{\text{+1}} \qw    &                   \qw\qwx&                   \qw\qwx&\gate{\text{+2}} \qw    &                   \qw\qwx&                   \qw\qwx&\gate{\text{+1}} \qw    &                   \qw\qwx&\gate{\text{+2}} \qw    &                   \qw\qwx&                   \qw\qwx&                   \qw\qwx&                   \qw\qwx &\rstick{\cdots} \qw \\
 &\lstick{\ket{0}}& \qw&\gate{{H}} \qw&  \push{\circled{0}} \qw\qwx&  \push{\circled{0}} \qw\qwx&  \push{\circled{0}} \qw\qwx&  \push{\circled{0}} \qw\qwx&  \push{\circled{0}} \qw\qwx&  \push{\circled{0}} \qw\qwx&  \push{\circled{0}} \qw\qwx&  \push{\circled{0}} \qw\qwx&  \push{\circled{0}} \qw\qwx&  \push{\circled{0}} \qw\qwx&  \push{\circled{0}} \qw\qwx&  \push{\circled{0}} \qw\qwx&  \push{\circled{0}} \qw\qwx&  \push{\circled{0}} \qw\qwx&  \push{\circled{0}} \qw\qwx&  \push{\circled{0}} \qw\qwx&  \push{\circled{0}} \qw\qwx&  \push{\circled{0}} \qw\qwx&  \push{\circled{0}} \qw\qwx&  \push{\circled{0}} \qw\qwx&  \push{\circled{0}} \qw\qwx&  \push{\circled{0}} \qw\qwx&  \push{\circled{0}} \qw\qwx&  \push{\circled{0}} \qw\qwx&  \push{\circled{0}} \qw\qwx&  \push{\circled{0}} \qw\qwx&  \push{\circled{0}} \qw\qwx&  \push{\circled{0}} \qw\qwx&  \push{\circled{0}} \qw\qwx&  \push{\circled{0}} \qw\qwx&  \push{\circled{0}} \qw\qwx&  \push{\circled{0}} \qw\qwx&  \push{\circled{0}} \qw\qwx&  \push{\circled{0}} \qw\qwx&  \push{\circled{0}} \qw\qwx&  \push{\circled{0}} \qw\qwx&  \push{\circled{0}} \qw\qwx&  \push{\circled{0}} \qw\qwx&  \push{\circled{0}} \qw\qwx&  \push{\circled{0}} \qw\qwx&  \push{\circled{1}}   \qw\qwx&  \push{\circled{1}}   \qw\qwx&  \push{\circled{1}}   \qw\qwx&  \push{\circled{1}}   \qw\qwx &\rstick{\cdots} \qw \\
 &\lstick{\ket{0}}& \qw&\gate{{H}} \qw&  \push{\circled{0}} \qw\qwx&  \push{\circled{0}} \qw\qwx&  \push{\circled{0}} \qw\qwx&  \push{\circled{0}} \qw\qwx&  \push{\circled{1}}   \qw\qwx&  \push{\circled{1}}   \qw\qwx&  \push{\circled{1}}   \qw\qwx&  \push{\circled{1}}   \qw\qwx&  \push{\circled{1}}   \qw\qwx&  \push{\circled{1}}   \qw\qwx&  \push{\circled{1}}   \qw\qwx&  \push{\circled{1}}   \qw\qwx&  \push{\circled{1}}   \qw\qwx&  \push{\circled{1}}   \qw\qwx&  \push{\circled{1}}   \qw\qwx&  \push{\circled{1}}   \qw\qwx&  \push{\circled{1}}   \qw\qwx&  \push{\circled{1}}   \qw\qwx&  \push{\circled{1}}   \qw\qwx&  \push{\circled{1}}   \qw\qwx&  \push{\circled{1}}   \qw\qwx&  \push{\circled{2}} \qw\qwx&  \push{\circled{2}} \qw\qwx&  \push{\circled{2}} \qw\qwx&  \push{\circled{2}} \qw\qwx&  \push{\circled{2}} \qw\qwx&  \push{\circled{2}} \qw\qwx&  \push{\circled{2}} \qw\qwx&  \push{\circled{2}} \qw\qwx&  \push{\circled{2}} \qw\qwx&  \push{\circled{2}} \qw\qwx&  \push{\circled{2}} \qw\qwx&  \push{\circled{2}} \qw\qwx&  \push{\circled{2}} \qw\qwx&  \push{\circled{2}} \qw\qwx&  \push{\circled{2}} \qw\qwx&  \push{\circled{2}} \qw\qwx&  \push{\circled{2}} \qw\qwx&  \push{\circled{2}} \qw\qwx&  \push{\circled{2}} \qw\qwx&  \push{\circled{0}} \qw\qwx&  \push{\circled{0}} \qw\qwx&  \push{\circled{0}} \qw\qwx&  \push{\circled{0}} \qw\qwx &\rstick{\cdots} \qw \\
 &\lstick{\ket{0}}& \qw&\gate{{H}} \qw&  \push{\circled{0}} \qw\qwx&  \push{\circled{1}}   \qw\qwx&  \push{\circled{2}} \qw\qwx&  \push{\circled{2}} \qw\qwx&  \push{\circled{0}} \qw\qwx&  \push{\circled{0}} \qw\qwx&  \push{\circled{0}} \qw\qwx&  \push{\circled{0}} \qw\qwx&  \push{\circled{0}} \qw\qwx&  \push{\circled{1}}   \qw\qwx&  \push{\circled{1}}   \qw\qwx&  \push{\circled{1}}   \qw\qwx&  \push{\circled{1}}   \qw\qwx&  \push{\circled{1}}   \qw\qwx&  \push{\circled{1}}   \qw\qwx&  \push{\circled{2}} \qw\qwx&  \push{\circled{2}} \qw\qwx&  \push{\circled{2}} \qw\qwx&  \push{\circled{2}} \qw\qwx&  \push{\circled{2}} \qw\qwx&  \push{\circled{2}} \qw\qwx&  \push{\circled{0}} \qw\qwx&  \push{\circled{0}} \qw\qwx&  \push{\circled{0}} \qw\qwx&  \push{\circled{0}} \qw\qwx&  \push{\circled{0}} \qw\qwx&  \push{\circled{0}} \qw\qwx&  \push{\circled{0}} \qw\qwx&  \push{\circled{0}} \qw\qwx&  \push{\circled{1}}   \qw\qwx&  \push{\circled{1}}   \qw\qwx&  \push{\circled{1}}   \qw\qwx&  \push{\circled{1}}   \qw\qwx&  \push{\circled{1}}   \qw\qwx&  \push{\circled{1}}   \qw\qwx&  \push{\circled{1}}   \qw\qwx&  \push{\circled{2}} \qw\qwx&  \push{\circled{2}} \qw\qwx&  \push{\circled{2}} \qw\qwx&  \push{\circled{2}} \qw\qwx&  \push{\circled{0}} \qw\qwx&  \push{\circled{0}} \qw\qwx&  \push{\circled{0}} \qw\qwx&  \push{\circled{0}} \qw\qwx &\rstick{\cdots} \qw \\
 &\lstick{\ket{0}}& \qw&\gate{{H}} \qw&  \push{\circled{1}}   \qw\qwx&  \push{\circled{2}} \qw\qwx&  \push{\circled{0}} \qw\qwx&  \push{\circled{2}} \qw\qwx&  \push{\circled{0}} \qw\qwx&  \push{\circled{0}} \qw\qwx&  \push{\circled{1}}   \qw\qwx&  \push{\circled{2}} \qw\qwx&  \push{\circled{2}} \qw\qwx&  \push{\circled{0}} \qw\qwx&  \push{\circled{0}} \qw\qwx&  \push{\circled{0}} \qw\qwx&  \push{\circled{1}}   \qw\qwx&  \push{\circled{1}}   \qw\qwx&  \push{\circled{2}} \qw\qwx&  \push{\circled{0}} \qw\qwx&  \push{\circled{1}}   \qw\qwx&  \push{\circled{1}}   \qw\qwx&  \push{\circled{1}}   \qw\qwx&  \push{\circled{2}} \qw\qwx&  \push{\circled{2}} \qw\qwx&  \push{\circled{0}} \qw\qwx&  \push{\circled{0}} \qw\qwx&  \push{\circled{0}} \qw\qwx&  \push{\circled{1}}   \qw\qwx&  \push{\circled{1}}   \qw\qwx&  \push{\circled{2}} \qw\qwx&  \push{\circled{2}} \qw\qwx&  \push{\circled{2}} \qw\qwx&  \push{\circled{0}} \qw\qwx&  \push{\circled{0}} \qw\qwx&  \push{\circled{0}} \qw\qwx&  \push{\circled{1}}   \qw\qwx&  \push{\circled{1}}   \qw\qwx&  \push{\circled{1}}   \qw\qwx&  \push{\circled{2}} \qw\qwx&  \push{\circled{1}}   \qw\qwx&  \push{\circled{1}}   \qw\qwx&  \push{\circled{2}} \qw\qwx&  \push{\circled{2}} \qw\qwx&  \push{\circled{0}} \qw\qwx&  \push{\circled{0}} \qw\qwx&  \push{\circled{1}}   \qw\qwx&  \push{\circled{1}}   \qw\qwx &\rstick{\cdots} \qw \\
 \\
}}
    \resizebox{\linewidth}{!}{\Qcircuit @R=1em @C=0.75em {
 \\
&\lstick{\cdots}&\gate{\text{+1}} \qw    &                   \qw    &                   \qw    &                   \qw    &                   \qw    &\gate{\text{+1}} \qw    &                   \qw    &                   \qw    &\gate{\text{+1}} \qw    &                   \qw    &                   \qw    &\gate{\text{+1}} \qw    &                   \qw    &                   \qw    &\gate{\text{+2}} \qw    &                   \qw    &                   \qw    &                   \qw    &                   \qw    &                   \qw    &                   \qw    &\gate{\text{+1}} \qw    &                   \qw    &                   \qw    &\gate{\text{+1}} \qw    &                   \qw    &\gate{\text{+2}} \qw    &                   \qw    &                   \qw    &                   \qw    &                   \qw    &\gate{\text{+2}} \qw    &                   \qw    &\gate{\text{+1}} \qw    &\gate{\text{+1}} \qw    &                   \qw    &                   \qw    &                   \qw    &\gate{\text{+2}} \qw    &                   \qw    &                   \qw    &                   \qw    &                   \qw    &                   \qw    &\gate{\text{+1}} \qw    &                   \qw    &                   \qw    &\gate{\text{+2}} \qw    &                   \qw    &                   \qw    &\gate{\text{+1}} \qw & \qwa\\
&\lstick{\cdots}&                   \qw\qwx&\gate{\text{+2}} \qw    &                   \qw    &\gate{\text{+1}} \qw    &                   \qw    &                   \qw\qwx&\gate{\text{+1}} \qw    &                   \qw    &                   \qw\qwx&\gate{\text{+2}} \qw    &                   \qw    &                   \qw\qwx&\gate{\text{+2}} \qw    &                   \qw    &                   \qw\qwx&\gate{\text{+1}} \qw    &                   \qw    &\gate{\text{+1}} \qw    &                   \qw    &                   \qw    &\gate{\text{+1}} \qw    &                   \qw\qwx&\gate{\text{+2}} \qw    &                   \qw    &                   \qw\qwx&                   \qw    &                   \qw\qwx&\gate{\text{+2}} \qw    &                   \qw    &\gate{\text{+2}} \qw    &                   \qw    &                   \qw\qwx&                   \qw    &                   \qw\qwx&                   \qw\qwx&\gate{\text{+1}} \qw    &                   \qw    &                   \qw    &                   \qw\qwx&\gate{\text{+1}} \qw    &                   \qw    &\gate{\text{+2}} \qw    &\gate{\text{+2}} \qw    &                   \qw    &                   \qw\qwx&\gate{\text{+2}} \qw    &                   \qw    &                   \qw\qwx&\gate{\text{+2}} \qw    &                   \qw    &                   \qw\qwx&     \qwa\\
 &\lstick{\cdots}&                   \qw\qwx&                   \qw\qwx&\gate{\text{+2}} \qw    &                   \qw\qwx&\gate{\text{+2}} \qw    &                   \qw\qwx&                   \qw\qwx&\gate{\text{+2}} \qw    &                   \qw\qwx&                   \qw\qwx&\gate{\text{+2}} \qw    &                   \qw\qwx&                   \qw\qwx&\gate{\text{+1}} \qw    &                   \qw\qwx&                   \qw\qwx&\gate{\text{+2}} \qw    &                   \qw\qwx&\gate{\text{+2}} \qw    &\gate{\text{+1}} \qw    &                   \qw\qwx&                   \qw\qwx&                   \qw\qwx&\gate{\text{+2}} \qw    &                   \qw\qwx&\gate{\text{+1}} \qw    &                   \qw\qwx&                   \qw\qwx&\gate{\text{+1}} \qw    &                   \qw\qwx&\gate{\text{+2}} \qw    &                   \qw\qwx&\gate{\text{+1}} \qw    &                   \qw\qwx&                   \qw\qwx&                   \qw\qwx&\gate{\text{+2}} \qw    &\gate{\text{+2}} \qw    &                   \qw\qwx&                   \qw\qwx&\gate{\text{+1}} \qw    &                   \qw\qwx&                   \qw\qwx&\gate{\text{+1}} \qw    &                   \qw\qwx&                   \qw\qwx&\gate{\text{+2}} \qw    &                   \qw\qwx&                   \qw\qwx&\gate{\text{+2}} \qw    &                   \qw\qwx&       \qwa\\
 &\lstick{\cdots}&\push{\circled{1}}   \qw\qwx&\push{\circled{1}}   \qw\qwx&\push{\circled{1}}   \qw\qwx&\push{\circled{1}}   \qw\qwx&\push{\circled{1}}   \qw\qwx&\push{\circled{1}}   \qw\qwx&\push{\circled{1}}   \qw\qwx&\push{\circled{1}}   \qw\qwx&\push{\circled{1}}   \qw\qwx&\push{\circled{1}}   \qw\qwx&\push{\circled{1}}   \qw\qwx&\push{\circled{1}}   \qw\qwx&\push{\circled{1}}   \qw\qwx&\push{\circled{1}}   \qw\qwx&\push{\circled{1}}   \qw\qwx&\push{\circled{1}}   \qw\qwx&\push{\circled{1}}   \qw\qwx&\push{\circled{1}}   \qw\qwx&\push{\circled{1}}   \qw\qwx&\push{\circled{1}}   \qw\qwx&\push{\circled{1}}   \qw\qwx&\push{\circled{1}}   \qw\qwx&\push{\circled{1}}   \qw\qwx&\push{\circled{1}}   \qw\qwx&\push{\circled{1}}   \qw\qwx&\push{\circled{1}}   \qw\qwx&\push{\circled{1}}   \qw\qwx&\push{\circled{1}}   \qw\qwx&\push{\circled{1}}   \qw\qwx&\push{\circled{1}}   \qw\qwx&\push{\circled{1}}   \qw\qwx&\push{\circled{1}}   \qw\qwx&\push{\circled{1}}   \qw\qwx&\push{\circled{1}}   \qw\qwx&\push{\circled{1}}   \qw\qwx&\push{\circled{1}}   \qw\qwx&\push{\circled{1}}   \qw\qwx&\push{\circled{1}}   \qw\qwx&\push{\circled{1}}   \qw\qwx&\push{\circled{1}}   \qw\qwx&\push{\circled{1}}   \qw\qwx&\push{\circled{1}}   \qw\qwx&\push{\circled{1}}   \qw\qwx&\push{\circled{1}}   \qw\qwx&\push{\circled{1}}   \qw\qwx&\push{\circled{1}}   \qw\qwx&\push{\circled{1}}   \qw\qwx&\push{\circled{1}}   \qw\qwx&\push{\circled{1}}   \qw\qwx&\push{\circled{1}}   \qw\qwx&\push{\circled{1}}   \qw\qwx&      \qwa\\
 &\lstick{\cdots}&\push{\circled{0}} \qw\qwx&\push{\circled{0}} \qw\qwx&\push{\circled{0}} \qw\qwx&\push{\circled{0}} \qw\qwx&\push{\circled{0}} \qw\qwx&\push{\circled{0}} \qw\qwx&\push{\circled{0}} \qw\qwx&\push{\circled{0}} \qw\qwx&\push{\circled{0}} \qw\qwx&\push{\circled{0}} \qw\qwx&\push{\circled{0}} \qw\qwx&\push{\circled{0}} \qw\qwx&\push{\circled{0}} \qw\qwx&\push{\circled{0}} \qw\qwx&\push{\circled{0}} \qw\qwx&\push{\circled{0}} \qw\qwx&\push{\circled{0}} \qw\qwx&\push{\circled{0}} \qw\qwx&\push{\circled{0}} \qw\qwx&\push{\circled{1}}   \qw\qwx&\push{\circled{1}}   \qw\qwx&\push{\circled{1}}   \qw\qwx&\push{\circled{1}}   \qw\qwx&\push{\circled{1}}   \qw\qwx&\push{\circled{1}}   \qw\qwx&\push{\circled{1}}   \qw\qwx&\push{\circled{1}}   \qw\qwx&\push{\circled{1}}   \qw\qwx&\push{\circled{1}}   \qw\qwx&\push{\circled{1}}   \qw\qwx&\push{\circled{1}}   \qw\qwx&\push{\circled{1}}   \qw\qwx&\push{\circled{1}}   \qw\qwx&\push{\circled{1}}   \qw\qwx&\push{\circled{2}} \qw\qwx&\push{\circled{2}} \qw\qwx&\push{\circled{2}} \qw\qwx&\push{\circled{2}} \qw\qwx&\push{\circled{2}} \qw\qwx&\push{\circled{2}} \qw\qwx&\push{\circled{2}} \qw\qwx&\push{\circled{2}} \qw\qwx&\push{\circled{2}} \qw\qwx&\push{\circled{2}} \qw\qwx&\push{\circled{2}} \qw\qwx&\push{\circled{2}} \qw\qwx&\push{\circled{2}} \qw\qwx&\push{\circled{2}} \qw\qwx&\push{\circled{2}} \qw\qwx&\push{\circled{2}} \qw\qwx&\push{\circled{2}} \qw\qwx&       \qwa\\
 &\lstick{\cdots}&\push{\circled{0}} \qw\qwx&\push{\circled{0}} \qw\qwx&\push{\circled{0}} \qw\qwx&\push{\circled{1}}   \qw\qwx&\push{\circled{1}}   \qw\qwx&\push{\circled{1}}   \qw\qwx&\push{\circled{1}}   \qw\qwx&\push{\circled{1}}   \qw\qwx&\push{\circled{1}}   \qw\qwx&\push{\circled{1}}   \qw\qwx&\push{\circled{1}}   \qw\qwx&\push{\circled{2}} \qw\qwx&\push{\circled{2}} \qw\qwx&\push{\circled{2}} \qw\qwx&\push{\circled{2}} \qw\qwx&\push{\circled{2}} \qw\qwx&\push{\circled{2}} \qw\qwx&\push{\circled{2}} \qw\qwx&\push{\circled{2}} \qw\qwx&\push{\circled{0}} \qw\qwx&\push{\circled{0}} \qw\qwx&\push{\circled{0}} \qw\qwx&\push{\circled{0}} \qw\qwx&\push{\circled{0}} \qw\qwx&\push{\circled{1}}   \qw\qwx&\push{\circled{1}}   \qw\qwx&\push{\circled{1}}   \qw\qwx&\push{\circled{1}}   \qw\qwx&\push{\circled{1}}   \qw\qwx&\push{\circled{1}}   \qw\qwx&\push{\circled{1}}   \qw\qwx&\push{\circled{2}} \qw\qwx&\push{\circled{2}} \qw\qwx&\push{\circled{2}} \qw\qwx&\push{\circled{0}} \qw\qwx&\push{\circled{0}} \qw\qwx&\push{\circled{0}} \qw\qwx&\push{\circled{0}} \qw\qwx&\push{\circled{0}} \qw\qwx&\push{\circled{0}} \qw\qwx&\push{\circled{0}} \qw\qwx&\push{\circled{1}}   \qw\qwx&\push{\circled{1}}   \qw\qwx&\push{\circled{1}}   \qw\qwx&\push{\circled{2}} \qw\qwx&\push{\circled{2}} \qw\qwx&\push{\circled{2}} \qw\qwx&\push{\circled{2}} \qw\qwx&\push{\circled{2}} \qw\qwx&\push{\circled{2}} \qw\qwx&\push{\circled{2}} \qw\qwx&       \qwa\\
 &\lstick{\cdots}&\push{\circled{2}} \qw\qwx&\push{\circled{2}} \qw\qwx&\push{\circled{2}} \qw\qwx&\push{\circled{0}} \qw\qwx&\push{\circled{0}} \qw\qwx&\push{\circled{1}}   \qw\qwx&\push{\circled{1}}   \qw\qwx&\push{\circled{1}}   \qw\qwx&\push{\circled{2}} \qw\qwx&\push{\circled{2}} \qw\qwx&\push{\circled{2}} \qw\qwx&\push{\circled{0}} \qw\qwx&\push{\circled{0}} \qw\qwx&\push{\circled{0}} \qw\qwx&\push{\circled{1}}   \qw\qwx&\push{\circled{1}}   \qw\qwx&\push{\circled{1}}   \qw\qwx&\push{\circled{2}} \qw\qwx&\push{\circled{2}} \qw\qwx&\push{\circled{0}} \qw\qwx&\push{\circled{1}}   \qw\qwx&\push{\circled{2}} \qw\qwx&\push{\circled{2}} \qw\qwx&\push{\circled{2}} \qw\qwx&\push{\circled{0}} \qw\qwx&\push{\circled{0}} \qw\qwx&\push{\circled{1}}   \qw\qwx&\push{\circled{1}}   \qw\qwx&\push{\circled{1}}   \qw\qwx&\push{\circled{2}} \qw\qwx&\push{\circled{2}} \qw\qwx&\push{\circled{0}} \qw\qwx&\push{\circled{1}}   \qw\qwx&\push{\circled{2}} \qw\qwx&\push{\circled{0}} \qw\qwx&\push{\circled{0}} \qw\qwx&\push{\circled{0}} \qw\qwx&\push{\circled{1}}   \qw\qwx&\push{\circled{2}} \qw\qwx&\push{\circled{2}} \qw\qwx&\push{\circled{2}} \qw\qwx&\push{\circled{1}}   \qw\qwx&\push{\circled{2}} \qw\qwx&\push{\circled{2}} \qw\qwx&\push{\circled{0}} \qw\qwx&\push{\circled{0}} \qw\qwx&\push{\circled{0}} \qw\qwx&\push{\circled{1}}   \qw\qwx&\push{\circled{1}}   \qw\qwx&\push{\circled{1}}   \qw\qwx&\push{\circled{2}} \qw\qwx&    \qwa\\
 \\
}}
    \caption{The circuit diagram for the QRCIQ representation for the image in Fig. \ref{fig:rgb}.}
    \label{fig:qrciq}
\end{sidewaysfigure*}
\end{rotatepage}
\fi

\vspace{12pt}
\end{document}